\documentclass{aa}
\usepackage{txfonts, graphicx, psfig, epsf, epsfig}

\newbox\grsign \setbox\grsign=\hbox{$>$}
\newdimen\grdimen \grdimen=\ht\grsign
\newbox\laxbox \newbox\gaxbox
\setbox\gaxbox=\hbox{\raise.5ex\hbox{$>$}\llap
        {\lower.5ex\hbox{$\sim$}}}\ht1=\grdimen\dp1=0pt
\setbox\laxbox=\hbox{\raise.5ex\hbox{$<$}\llap
        {\lower.5ex\hbox{$\sim$}}}\ht2=\grdimen\dp2=0pt
\newcommand{\simlt}{\mathrel{\copy\laxbox}}
\newcommand{\simgt}{\mathrel{\copy\gaxbox}} 

\begin{document}
   \title{Properties of discrete X-ray sources \\ 
          in the starburst spiral galaxy M\,83}


   \author{Roberto Soria \inst{1} 
          \and
          Kinwah Wu \inst{1}
          }

   \offprints{R.~Soria ({\tt rs1@mssl.ucl.ac.uk})}

   \institute{Mullard Space Science Laboratory, 
          University College London, Holmbury St Mary, 
          Surrey RH5 6NT, UK  \\
              emails: {\tt{Roberto.Soria@mssl.ucl.ac.uk}}, {\tt{kw@mssl.ucl.ac.uk}} }

   \date{Received December 2002; accepted July 2003  }

   \abstract{
We have identified 127 discrete sources 
  in a {\it Chandra} ACIS observation of M\,83, 
  with a detection limit of $\approx 3 \times 10^{36}$ erg s$^{-1}$ 
  in the $0.3$--$8.0$ keV band.  
We discuss the individual X-ray spectral and time-variability properties 
  of $\approx 20$ bright sources with luminosities 
	$\simgt 10^{38}$ erg s$^{-1}$,  
  and the statistical properties of the whole sample.
About one third of the bright sources show X-ray spectra 
with a blackbody component at temperatures $\simlt 1$ keV, 
plus a powerlaw component with $\Gamma \approx 2.5$, 
  typical of X-ray binaries in a soft state; 
  another third have powerlaw spectra with $\Gamma \approx 1.5$,  
  consistent with X-ray binaries in a hard state. 
Two bright sources show emission lines on a hard powerlaw continuum, 
and are probably X-ray binaries surrounded by a photo-ionized nebula or stellar wind. 
Among the other bright sources, we also identified two supernova remnant 
candidates, 
with optically-thin thermal plasma spectra at temperatures $\sim 0.5$ keV. 
The two brightest supersoft sources have blackbody temperatures 
  $kT \approx 70$ eV and luminosities $\sim 10^{38}$ erg s$^{-1}$.
Two candidate X-ray pulsars are detected with periods $\approx 200$~s.
One X-ray source corresponds to the core of a background 
FRII radio galaxy.
The discrete sources can be divided into three groups, 
  based on their spatial, color and luminosity distributions. 
The first group comprises supersoft sources with no detected 
emission above 1 keV 
  and blackbody spectra at temperatures $< 100$ eV. 
The second group consists of soft sources 
  with little or no detected emission above 2 keV. 
They are strongly correlated with H$\alpha$ emission 
  in the spiral arms and starburst nucleus, tracing 
a young population. Their relative abundance  
depends on the current level of star-forming activity 
in the galaxy. Most of them are likely to be supernova remnants. 
The sources in the third group are mostly X-ray binaries, 
reaching higher X-ray luminosities than sources in the other 
two groups. Being a mixture of old low-mass and young high-mass systems, 
the whole group appears to be of intermediate age when correlated 
with the H$\alpha$ emission. The color-color diagrams allow us 
to distinguish between sources in a soft and hard state.
}

\authorrunning{R. Soria \& K. Wu}
   \maketitle
%

\section{Introduction}  

M\,83 (NGC 5236) is a nearby spiral galaxy (Hubble type SAB(s)c)  
  with a circum-nuclear starburst.   
It is oriented almost face-on, with an inclination $i=24^{\circ}$ 
  (Talbot et al. 1979).  
Its distance was estimated to be 8.9~Mpc 
  by Sandage \& Tamman (1987),  
  but a value of 3.7~Mpc was obtained 
  more recently by de Vaucouleurs et al.\ (1991).   
A distance $\simlt 5$ Mpc would place M\,83 in the Centaurus\,A group, 
  whose galaxies have high velocities and a large spread in morphology 
  (de Vaucoulers 1979; C{\^ o}t{\' e} et al.\ 1997), 
  suggesting that the group is not yet virialized, 
  and that merging and tidal interactions may be frequent occurrences 
  among its members. 
Here and hereafter, we adopt the distance estimate of 3.7~Mpc.

                            
\begin{figure*}[t] 
\begin{center} 
\epsfig{figure=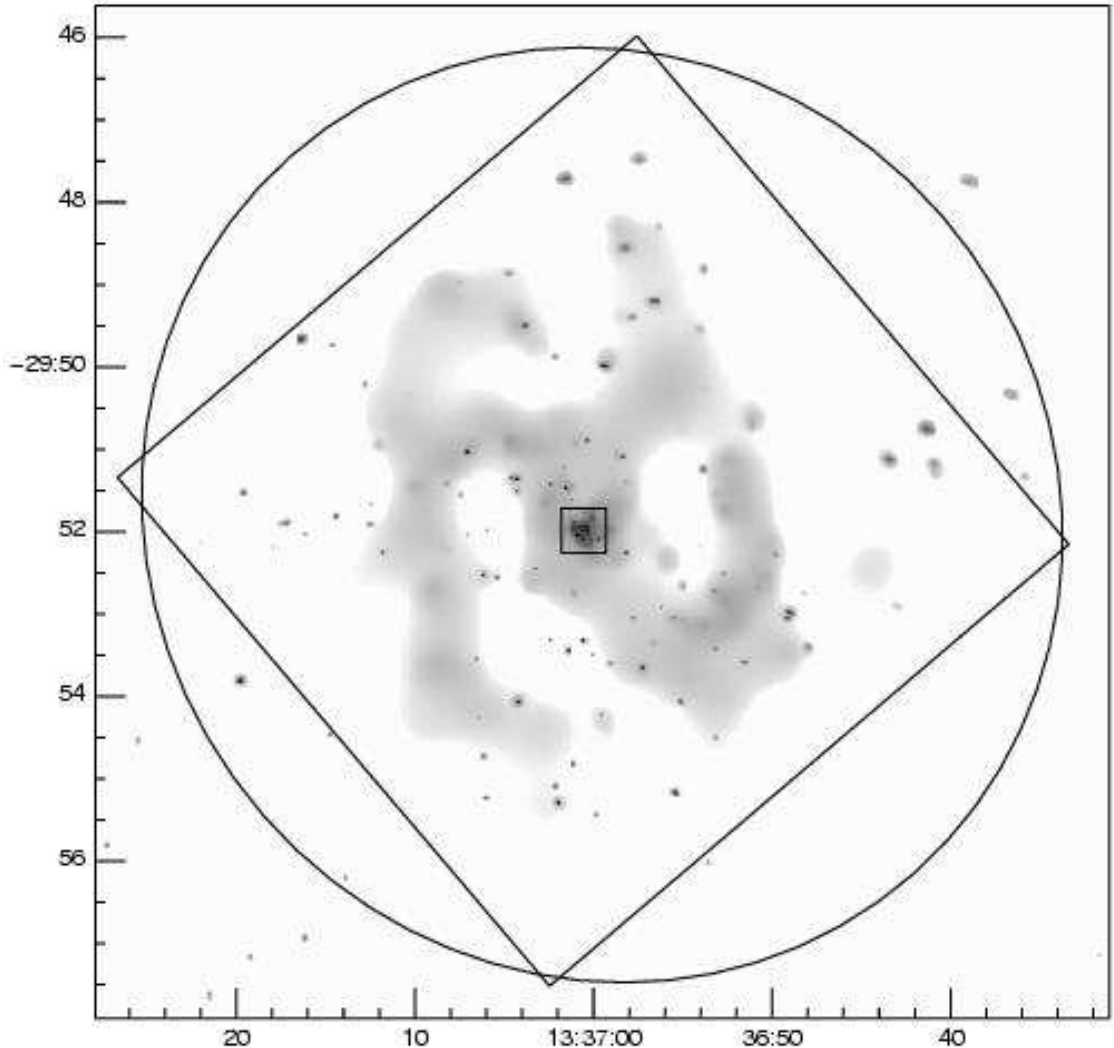,width=16.0cm}
\end{center}
\caption{
  Adaptively-smoothed greyscale {\it Chandra} ACIS image of M\,83, 
  in the $0.3$--$8.0$ keV band (square-root scale, arbitrary 
  zeropoint). 
  The $D_{25}$ ellipse and the S3 chip boundaries are overplotted. 
  The region inside the central square is shown 
in greater detail in Fig.~\ref{fig:imagegrey2}.}
\label{fig:imagegrey}
\end{figure*} 


                            
\begin{figure}[t] 
\begin{center} 
\epsfig{figure=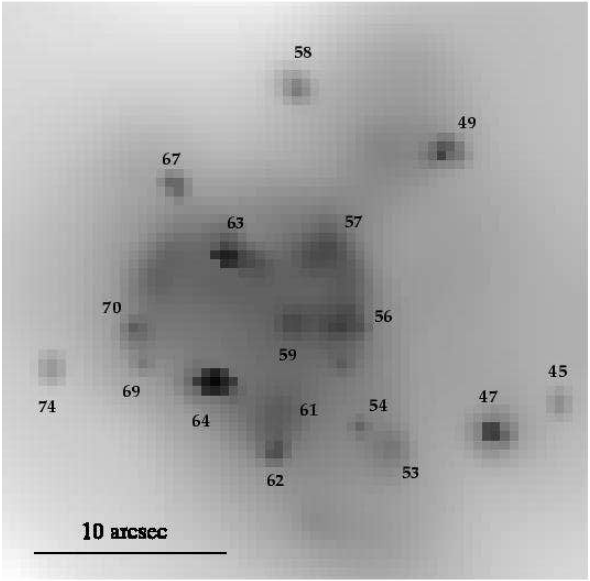,width=8.0cm, angle=270}
\end{center}
\caption{
  Close-up view of the starburst nucleus. The source numbers 
refer to Table A.1. North is up, and East is left. }
\label{fig:imagegrey2}
\end{figure} 


                            
\begin{figure}[t] 
\begin{center} 
\epsfig{figure=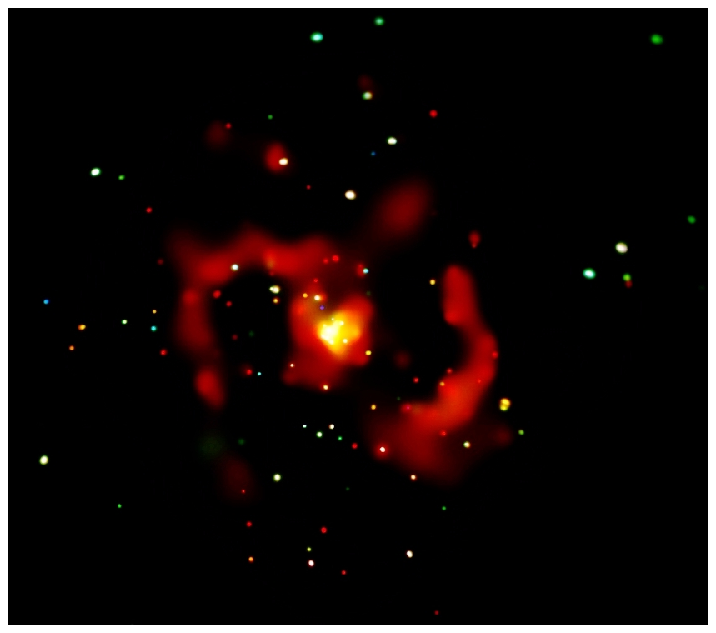,width=8.0cm} 
\end{center}
\caption{
   A ``true-colour'' {\it Chandra} ACIS image of M\,83  
     shows about 130 discrete sources 
     (down to a luminosity of $\approx 3 \times 10^{36}$ erg s$^{-1}$ 
     for the assumed distance of 3.7 Mpc) 
     and diffuse emission in the starburst nucleus and along the spiral arms. 
  The colors are: red for $0.3$--$1.0$~keV; 
                green for $1.0$--$2.0$~keV; 
                blue for $2.0$--$8.0$~keV.   
  The size of the image is $12\arcmin \times 10\arcmin$. 
  North is up, East is left.  
}
\label{fig:image}
\end{figure}


M\,83 has two grand-design spiral arms with ongoing star formation 
  (eg, Talbot et al.\ 1979; Tilanus \& Allen 1993).  
In addition, it has an even more active nuclear region  
  ($\simlt 10$\arcsec~in radius)  
  which is currently undergoing a violent starburst  
  (Gallais et al.\ 1991; Elmegreen et al. 1998; Harris et al.\ 2001).  
The star formation rate due to the circum-nuclear starburst   
  is estimated to be $\approx 0.1-0.2$~M$_\odot$~yr$^{-1}$ 
  (de Vaucouleurs et al.~1983; Harris et al.\ 2001), 
  similar to the rate of star formation in the rest of the galaxy 
  (eg, Buat et al.\ 2002).    
It was suggested that intense star formation may have been   
  induced by the last close encounter with the dwarf S0 NGC~5253  
  (Bohlin et al.\ 1983; van den Bergh 1980). 
   
M\,83, together with M\,82, NGC~253 and NGC~4395, 
  accounts for about 25\% of today's total rate of star formation  
  within a distance of 10~Mpc from the Milky Way (Heckman 1997).  
Massive stars are progenitors of supernova remnants (SNRs) and X-ray binaries (XRBs), 
  the brightest non-nuclear X-ray sources in galaxies.     
Observations have shown that the X-ray properties of normal galaxies  
  are closely related to their recent star formation activity.  
Galaxies with active star formation 
  (eg,\ M\,82, Matsumoto et al.\ 2001; NGC~4038/4039, Fabbiano et al.~2001)
  have a large population of bright discrete sources, 
  but quiescent galaxies (eg,\ M\,31, Supper et al.\ 2001; Shirey et al.\ 2001)
  are often deficient in very bright sources. 
Moreover, the spatial distribution of the sources 
  is different in different galactic components. 
In the spiral galaxy M\,81 
  the galactic disk contains more bright sources than the bulge 
  (e.g.\ Tennant et al.\ 2001; Swartz et al.\ 2002, 2003), 
  and the brighter sources in the disk 
  tend to be located closer to the spiral arms (Swartz et al.\ 2003).  
For these reasons, the X-ray source population is a probe  
  of the recent star formation activity and dynamics of the host galaxy  
  (Wu 2001; Wu et al.\ 2003; Prestwich 2001; Prestwich et al.\ 2003), 
  and high spatial resolution X-ray observations 
  of M\,83 are particularly important in this context. 

M\,83 was observed in the X-ray band by {\em Chandra} on 2000 April 29.
The galaxy had also been observed 
  by {\em Einstein} in 1979--1981 (Trinchieri et al.~1985),  
  by {\em GINGA} in 1988 (Ohashi et al.\ 1990), 
  by {\em ROSAT} in 1992--1994 
  (PSPC, Ehle et al.\ 1998; HRI, Immler et al.\ 1999),  
  and by {\em ASCA} in 1994 (Okada et al.~1997). 
Thirteen ``discrete'' sources were found in the {\em ROSAT} PSPC image, 
  and 37 with the HRI. 
The starburst nuclear region was unresolved in the {\em ROSAT} images.    
Our preliminary analysis of the {\em Chandra} data 
  identified a total of 81 sources detected with signal-to-noise ratio $> 3.5$, 
  and resolved the {\em ROSAT} nuclear source into at least 15 discrete 
  sources embedded in strong diffuse emission (Soria \& Wu 2002).  
Here we report more comprehensive results of our individual and statistical  
  analysis of the discrete source population.    
We compile a more complete list of the sources detected in the ACIS-S3 chip, 
  and determine the spectral properties of the brightest individual objects. 
We also quantify several statistical properties of the discrete sources  
  and compare some of them with those found for the M\,81 source population.  
  
We organize the paper as follows.  
In Sect.~2 we describe the general source extraction and data analysis procedures; 
  in Sect.~3 and 4 we present the results of our spectral and timing analysis 
  of the brightest sources.  
The statistical properties of the discrete source population  
  and the comparison with M\,81 are discussed in Sect.~5.  

\section{Source detection} 
\label{identification} 

The {\em Chandra} observation analyzed here was carried out 
  on 2000 April 29 (ObsID: 793), 
  with the ACIS-S3 chip at the focus. 
We retrieved the data from the {\it Chandra} X-ray Center archives.   
The total exposure time was 50.978~ks; 
  after screening out observational intervals with strong background flares, 
  we retained a good time interval of 50.851~ks (using less stringent 
  background rejection criteria compared to Soria \& Wu (2002)).   

We analyzed the data of the S3 chip  
  with the CXC software {\small{CIAO}} version 2.2.1. 
The source-finding routine {\it wavdetect} was used 
  to identify the discrete sources 
  in the full ($0.3$--$8.0$ keV) band,    
  and we calculated an exposure map at $1.0$~keV 
  for correcting the net count rates.  
We also compared the lists of sources 
  obtained from the two routines {\it wavdetect} and {\it celldetect} 
  (the latter was used in Soria \& Wu 2002) and found only very small  
  differences. We chose to adopt here the positions and count rates 
  obtained with {\it wavdetect}.
We first divided the $0.3$--$8.0$~keV band into three narrower bands:  
  the ``soft'' ($0.3$--$1.0$~keV) band, 
  the ``medium'' ($1.0$--$2.0$~keV) band and 
  the ``hard'' ($2.0$--$8.0$ keV) band.   
We then produced Filtered Event Files for the three energy bands  
  and applied {\it wavdetect} to each narrow-band image  
  using exposure maps at 0.7, 1.5 and 4.0~keV respectively.
Each {\it wavdetect} run produced a source list with slightly different 
positions: we examined and compared the position of each source 
in the four lists to identify the corresponding objects. 
In all cases in which the same source was detected in more than one band, 
the positional difference was always $< 1\arcsec$. We listed  
the coordinates obtained from the full band in Table A.1 
(uncertainty of $\approx 0\farcs5$). In the same Table, 
we also listed the net counts in the full, 
soft, medium and hard bands. 

We identified 127 discrete sources in the S3 chip  
with a signal to noise ratio $S/N> 3$ in the full band.
Some of these sources are not found by {\it wavdetect} 
in one or more narrow bands, even when we lower the detection 
threshold: in those cases, the net counts were estimated 
from an individual spectral analysis of each source (spectra 
extracted with {\it psextract}). 
When the net count values are given in brackets in Table A.1, 
the source has $S/N < 3$ in that band.

The 90\% uncertainty in the {\em Chandra} positions is $0\farcs{6}$.  
To improve the absolute positions of the sources, 
  we looked for cross-correlations with sources 
  in the International Celestial Reference Frame (Ma et al.\ 1998), 
  the Tycho~2 catalogue, the 2MASS survey, 
  the {\em Hubble Space Telescope} ({\em HST}) Guide Star Catalogue 
  and the US Naval Observatory Precision Measuring Machine Catalogue. 
However, none of those sources could be identified with sources 
in the {\it Chandra} ACIS-S3 image. 
We also compared the position of the nuclear X-ray source 
  with the optical/UV position of the IR nucleus 
  deduced from the {\em HST} WFPC2 archival observations.   
We found that 
  a shift of $0\farcs{5}$ to the {\em Chandra} coordinates   
  could align the X-ray and optical/UV positions of the nucleus.  
However, the WFPC2 coordinates themselves have also 
uncertainties of $0\farcs{5}$.    

We used the results of the Chandra 1Ms Deep Field observations 
(Rosati et al.\ 2002) to estimate the fraction 
of background sources. We obtained that $20\pm 4$ of our sources 
($\approx 15$\%) should be background AGN. The background 
contribution to sources brighter than 50 total counts 
($\approx 0.001$ cts s$^{-1}$) is expected to be $7 \pm 3$;
to sources brighter than 100 total counts 
($\approx 0.002$ cts s$^{-1}$): $3 \pm 1$; 
to sources brighter than 200 total counts 
($\approx 0.004$ cts s$^{-1}$): $1 \pm 0.5$.

Finally, we obtained a smoothed, grayscale X-ray image, plotted in 
Fig.~\ref{fig:imagegrey} with its coordinate grid, the $D_{25}$ ellipse 
and the S3 chip boundary\footnote{This image includes the most 
luminous source inside the 
$D_{25}$ ellipse, located at {R.A.~(2000) $=$ 13$^h$\,37$^m$\,$19\farcs{8}$}, 
{Dec.~(2000) $=$ $-$29$^{\circ}$\,53\arcmin\,49\arcsec}, ie, $\approx 4\farcm{6}$ 
south-east of the nucleus. It has a net count rate 
of 0.048 cts s$^{-1}$ and a luminosity $\approx 10^{39}$~erg~s$^{-1}$. 
However, it is located in chip S2 (just outside the S3 chip boundary), 
and therefore is not discussed in this paper. See Soria \& Wu (2002).}. 
A close-up view of the starburst nuclear region is shown in 
Fig.~\ref{fig:imagegrey2}. We also show a ``true-color'' 
image (Fig.~\ref{fig:image}), obtained from adaptively-smoothed 
images in the soft, medium and hard bands (CIAO task {\it csmooth} 
with minimal significance of the signal under the Gaussian 
kernel $= 3$ and maximal significance $=5$).

\section{Spectral properties of individual sources}  
\label{spectra}  

We modelled the spectra with higher $S/N$: about twenty sources 
are detected with $\simgt 300$~counts in the $0.3-8.0$~keV band, 
and two other bright, soft sources have fluxes $>100$~counts 
in the $0.3-1.0$~keV band alone. 
We also modelled the spectra of two fainter  
($\sim 150$ counts) sources because of their 
interesting physical nature (an X-ray pulsar 
and a background radio galaxy, see Sect.~3.5 and 4).
For sources that are not bright enough for meaningful 
spectral modelling, we only determined their colors and color ratios.   

We fixed the redshift $z = 0.00172$ and the foreground Galactic absorption 
column density $n_{\rm H} = 4.0 \times 10^{20}$~cm$^{-2}$ 
  (Predehl \& Schmitt 1995; Schlegel et al.~1998), 
  but we left the hydrogen column density component within M\,83 
  as a free parameter\footnote{We were also aware 
of the time-dependent degradation of the ACIS Quantum Efficiency 
at soft energies. However, we checked that this effect is still negligible 
for our 2000 April observation.}. 
We first considered simple one-component models in {\footnotesize XSPEC} 
(Arnaud 1996): 
  powerlaw, blackbody, disk-blackbody and Raymond-Smith  
  (for optically-thin thermal plasma). In most cases, 
they gave acceptable fits; however, a few sources required 
more complex spectral models. For sources suspected to be accreting 
compact objects, we also considered the comptonization model bmc.
Spectra showing emission-line features 
  (from candidate SNRs or wind-accreting XRBs),  
  were also fitted with Gaussian lines in addition 
to a smooth continuum.
  
In total, we modelled the spectra of 23 bright sources.
For convenience in our discussion, 
  we have grouped them according to their general spectral 
properties and possible physical interpretation.          
A sample of the fitted spectra are shown in this Section, 
  the others are shown in Appendix B.
The fit parameters of all 23 sources are tabulated in Appendix B. 

\subsection{Nuclear source}   
\label{nuclear}    

                            
\begin{figure}
\begin{center} 
\vspace*{0.35cm}
\epsfig{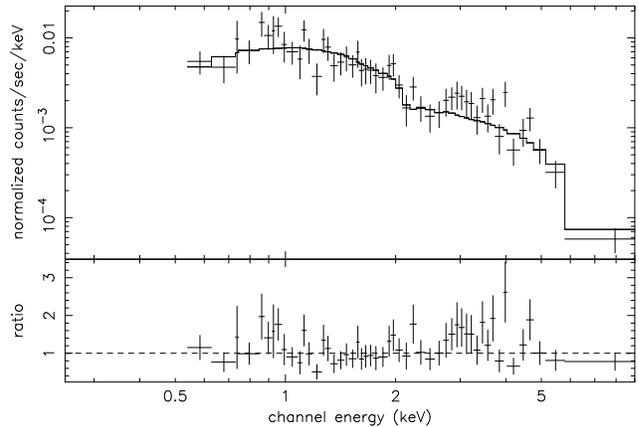}
\end{center}
\caption{
 Background subtracted X-ray spectral distribution (crosses), 
together with the
   best fitting absorbed power law model (histogram, upper panel), 
for source No.~63, coincident with the optical nucleus of M\,83.
 The ratio of data versus model is plotted in the lower panel. 
Error bars along  
the Y axis show statistical uncertainties, those along the X axis 
represent the width of the energy bins.
The X-ray spectrum is well fitted by an absorbed powerlaw 
    with $\Gamma \approx 1.45$ (fit parameters in Table B.1). }
\label{fig:nuclear_fit}
\end{figure}  


The presence of two galactic nuclei, 
  each with a mass of about $1.3 \times 10^7$~M$_\odot$, 
  was inferred from stellar kinematic data (Thatte et al.~2000). 
One is a strong IR and optical source; 
  the other is not visible at either wavelength.  
We have identified an X-ray point source (No.~63, see Appendix A) 
  with a position almost coincident with that of the IR/optical nucleus  
  (Soria \& Wu 2002). 
We consider this X-ray source as the galactic nucleus,    
  though we cannot rule out the possibilities  
  of a stellar-mass X-ray source observed 
  near the galactic center by chance.    

The nuclear source is embedded in strong, inhomogeneous 
diffuse X-ray emission, 
which makes it difficult to subtract the background contribution, 
especially at low energies. However, its X-ray spectrum 
is well fitted by a single powerlaw (Fig.~\ref{fig:nuclear_fit}), 
with no need for an additional thermal component.
We obtain a powerlaw photon index $\Gamma = 1.45^{+0.16}_{-0.24}$  
  and an intrinsic absorption column density 
$n_{\rm H} = (1.25^{+0.80}_{-0.93}) \times 10^{21}$~cm$^{-2}$, 
  with $\chi^2_\nu = 1.06$ for 49~d.o.f. (Table B.1). 
This is consistent, within the errors, with the result 
of our preliminary analysis in Soria \& Wu (2002). 
The emitted X-ray luminosity  
  is $L_{\rm x} \approx 2.3 \times 10^{38}$~erg~s$^{-1}$ 
  in the $0.3$--$8.0$~keV band.

The powerlaw spectrum of the nuclear source 
  is similar to those of supermassive black holes (BHs)
  in active galaxies ($\Gamma \approx 1.5$).  
The absence of a blackbody component in the $0.3$--$8.0$~keV band 
  is consistent with a BH mass $\simgt 10^7$~M$_\odot$, 
  implying that the inner accretion disk 
  has an effective blackbody temperature below 0.1~keV.   
The unabsorbed X-ray luminosity deduced from the fitted spectrum 
  is comparable to the luminosity of stellar-mass accreting objects, 
almost ten million times below the Eddington limit of the galactic nucleus:  
if the source is indeed a supermassive BH,
its current accretion rate must be extremely low.

\subsection{Supersoft sources} 
\label{supersoft}     

                          
\begin{figure}
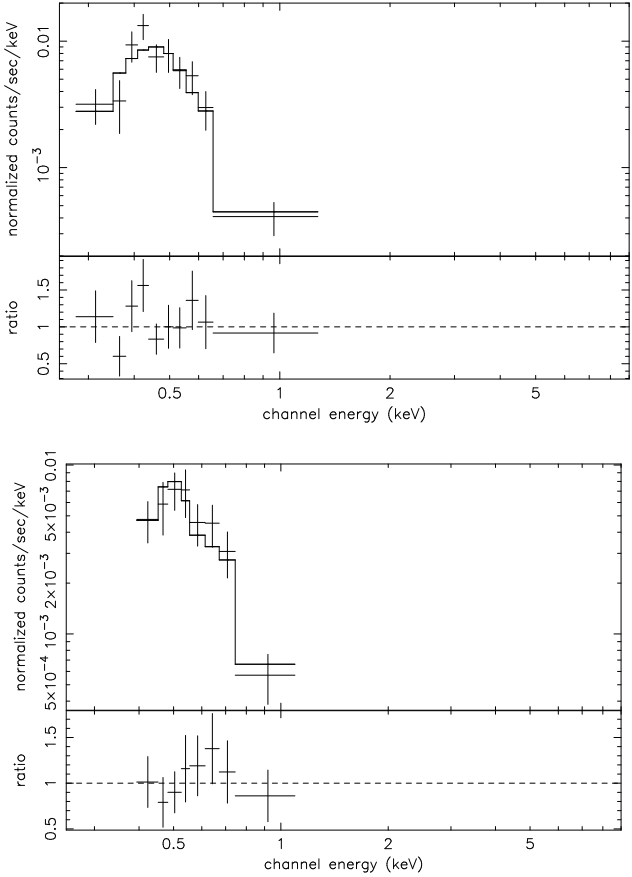

\begin{center} 
\vspace*{0.35cm}
\epsfig{figure=ss1n.ps,width=5.6cm,angle=270} \\ 
\vspace*{0.35cm}
\epsfig{figure=ss2n.ps,width=5.6cm,angle=270} 
\end{center}
\caption{
  The spectra of the two brightest supersoft sources (top: No.~68; 
bottom: No.~96); see Table B.2 
for the fit parameters.}
\label{fig:ss_fit}
\end{figure}  


Two bright sources, No.~68 and 96, have detectable emission only below 1~keV, 
and can be classified as ``supersoft''.
Neither was detected by {\it ROSAT} because of insufficient photon counts.  
We tried fitting their spectra with an absorbed blackbody, powerlaw 
and optically-thin thermal plasma models.
The powerlaw model is inconsistent with the data, 
and the optically-thin plasma model requires unphysically 
low abundances. Both spectra are instead well fitted 
by a blackbody model (Fig.~\ref{fig:ss_fit} and Table B.2).    
The fitted temperatures are $kT_{\rm bb} = 65^{+14}_{-13}$~eV
for source No.~68, and $kT_{\rm bb} = 58^{+82}_{-28}$~eV 
for source No.~96.

 
Most supersoft sources are believed to be 
  accreting white dwarfs undergoing (quasi-)steady nuclear burning 
  on their surface (van den Heuvel et al.\ 1992; Rappaport et al.~1994).   
Other models for supersoft sources 
  include SNRs, accreting neutron stars (NSs) with large photospheres, 
  intermediate-mass BHs, symbiotic systems, 
  hot cores of young planetary nebulae 
  and stripped core of tidally disrupted stars 
  (eg, Di\,Stefano \& Kong 2003).   
All these systems can have blackbody-like spectra 
  with characteristic temperatures $\approx 50$--$80$~eV.
The parameters of the blackbody model for sources No.~68 and 96 
  are also similar to those of the supersoft sources in the nearby spirals   
  M\,31 (Kahabka 1999) and M\,81 (Swartz et al.\ 2002).   

    
\subsection{Candidate SNRs and emission-line sources}  
\label{supernova}   

                          
\begin{figure}
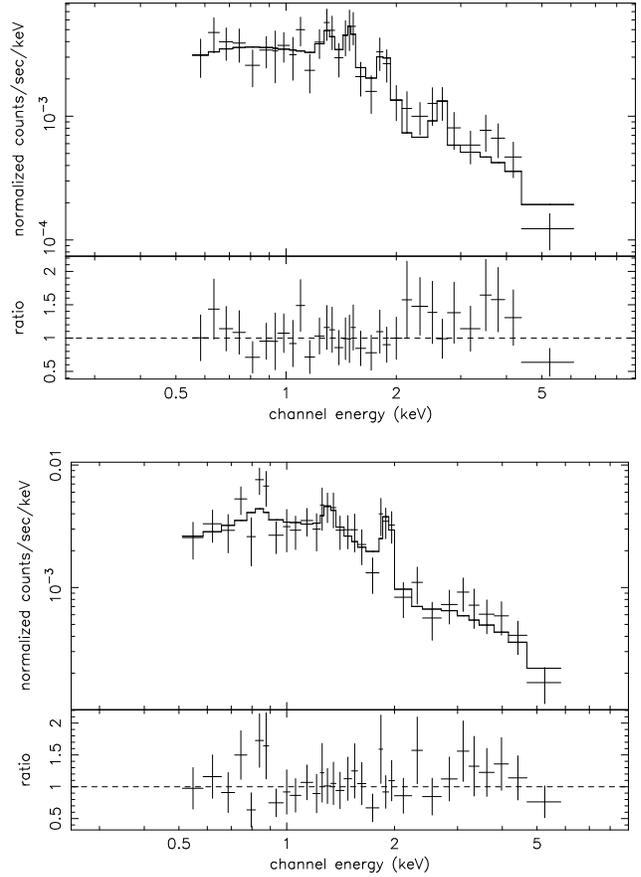

\begin{center} 
\vspace*{0.35cm}
\epsfig{figure=s03_line.ps,width=5.6cm,angle=270} \\
\vspace*{0.35cm} 
\epsfig{figure=s27_line.ps,width=5.6cm,angle=270}
\end{center}
\caption{ 
Spectra of sources No.~3 (top) and 27 (bottom) 
together with a best-fitting absorbed 
power law model with emission lines (see Table B.3).}
\label{fig:sn_line_fit}
\end{figure} 

 
                          
\begin{figure}
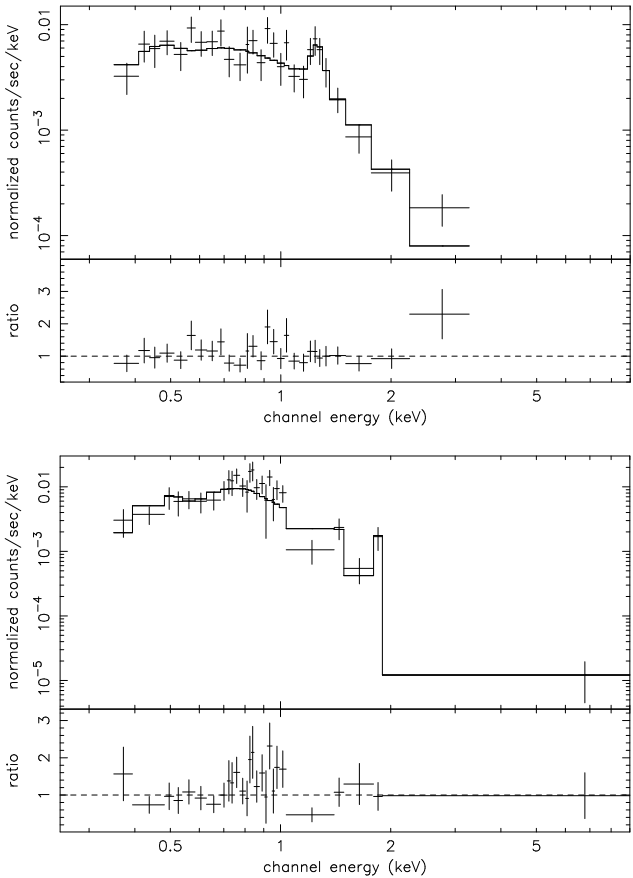

\begin{center} 
\vspace*{0.35cm}
\epsfig{figure=s08_line.ps,width=5.6cm,angle=270} \\
\vspace*{0.35cm}
\epsfig{figure=s56_line.ps,width=5.6cm,angle=270}
\end{center}
\caption{ 
Spectra of sources No.~8 (top) and 56 (bottom) 
together with a best-fitting single-temperature thermal plasma model 
(see Table B.3). }
\label{fig:sn1968_fit}
\end{figure} 

 
                         
\begin{figure}
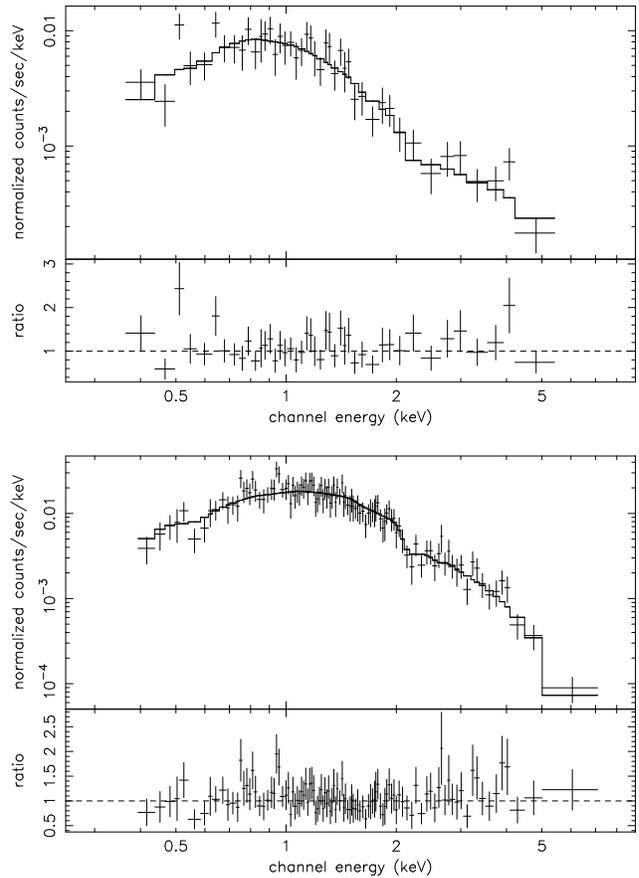

\begin{center} 
\vspace*{0.35cm} 
\epsfig{figure=s73.ps,width=5.6cm,angle=270} \\ 
\vspace*{0.35cm} 
\epsfig{figure=s86.ps,width=5.6cm,angle=270}  
\end{center}
\caption{ 
Spectra of sources No.~73 (top) and 86 (bottom) 
together with a best-fitting absorbed 
power law model (see Table B.4).}
\label{fig:sn_7386_fit}
\end{figure} 


The spectra of sources No.~3, 8, 27 and 56 show line-like features, 
  suggesting the possibility of emission from optically-thin thermal plasma. 
Fitting their spectra with a powerlaw model (Table B.3) shows 
  that the underlying continuum is very soft for two of them 
  (No.~8 and 56 require an unphysically high photon index $\Gamma > 3.3$) 
  and harder for the other two ($\Gamma \approx 1.5$). 
A bremsstrahlung model (Table B.3) gives 
  temperatures $\simlt 1$ keV for the two soft sources, 
  and temperatures $\simgt 10$ keV for the two hard ones.
The same temperature range is obtained from a Raymond-Smith 
model. For the two softer sources, the lack 
of a strong Fe L-line complex requires a low metal abundance; 
on the other hand, this underestimates the Mg and Si emission.
The S/N is too low to allow for a meaningful fitting of different 
metal abundances for different elements.

Finally, we tried to model the spectra of these four sources 
  with gaussian lines added to a smooth continuum: 
  we chose a powerlaw continuum for the hard sources No.~3 and 27, 
  and a bremsstrahlung continuum for the soft sources No.~8 and 56.
Four emission lines are significantly detected in source No.~3 (Table B.3): 
  one at $E = 1.32^{+0.04}_{-0.05}$ keV, 
  consistent with Mg\,{\footnotesize XI} (1.33 keV); 
  one at $E = 1.51^{+0.04}_{-0.03}$ keV, 
  suspected to be Mg\,{\footnotesize XII} (1.47 keV); 
  one at $E = 1.85^{+0.04}_{-0.03}$ keV, 
  consistent with the Si\,{\footnotesize XIII} triplet (1.84--1.87 keV); 
  and one at $E = 2.60^{+0.19}_{-0.10}$ keV, 
  possibly due to S\,{\footnotesize XVI} (2.62 keV). 
Taken together, 
  these four lines require an ionisation parameter $\log \xi \approx 2.5$ 
  for the emitting plasma.
Some of these metal lines are also seen in sources No.~8, 27 and 56.
The spectral modelling and luminosity estimate for source No.~56 
has the largest uncertainty, 
because the source is located in the nuclear starburst ring, and 
it is embedded in strong, non-uniform diffuse emission 
from hot thermal plasma. Its spectrum may also be contaminated 
by other nearby, unresolved sources.

Thermal plasma temperatures $\approx 0.7$ keV
and the presence of emission lines from $\alpha$-elements  
make sources No.~8 and 56 strong SNR candidates; however, 
long-term time-variability studies 
are required to confirm this identification\footnote{After completing 
our spectral analysis, the {\it Chandra} ACIS data of a short (good time 
interval of 6.4 ks) 
observation of M\,83 from 2001 April have become available in the public archive. 
A quick-look analysis of the new dataset suggests that the flux 
of source No.~8 has increased by a factor of 2. If this is confirmed, 
the SNR identification becomes less likely, and the thermal plasma 
is more likely to be ionized by an accreting XRB. A detailed comparison 
of all the sources' fluxes and colors in the two observations, and between 
the {\it ROSAT} and {\it Chandra} observations, is beyond 
the scope of this paper.}. 

The harder spectrum seen from sources No.~3 and 27 
requires much higher plasma temperatures, or a powerlaw continuum. 
We cannot rule out an SNR origin for these two sources: 
in young SNRs, the forward shock 
(propagating into the circumstellar medium) produces hard X-ray 
emission, while the reverse shock (propagating into the ejecta) 
produces soft X-ray emission. If the cooling shell at the contact 
discontinuity is optically thick, the soft component 
is not detectable (Chevalier \& Fransson 1994).
High-temperature optically-thin thermal emission  
  has been detected in some young Galactic remnants 
  (eg, SN~1998S, Pooley et al.\ 2002). 

However, we note that a hard powerlaw continuum with the superposition 
of emission lines  
can originate from X-ray binaries surrounded by a photo-ionized nebula.  
For example, strong recombination lines are seen in the Galactic high-mass XRBs 
Cyg X-3, Cen X-3 and Vela X-1 (eg, Liedahl et al.\ 2000 
and references therein), which are accreting from highly ionized stellar winds, 
and reach X-ray luminosities $\sim 10^{38}$ erg s$^{-1}$.  
(In Vela X-1, the differential emission measure suggests 
an ionization parameter $2 \simlt \log \xi \simlt 3$ 
similar to what is observed in source No.~3; however, 
the absorbing column density seen by the continuum emission 
is higher in Vela X-1; see Sako et al.\ 1999).
We suggest that source No.~3 and 27 may be such accreting systems in M\,83.

No radio or optical counterparts were found 
  for sources No.~3, 8 and 27, 
  nor do they correspond to any of the six historical supernovae (SNe) 
in the galaxy\footnote{http://cfa-www.harvard.edu/cfa/ps/lists/Supernovae.html}. 
Conversely, no X-ray sources are detected at the positions of the five   
historical SNe observed in the disk. However, one of the historical SNe, 
Type-II SN 1968L (Thackeray 1968; Wood \& Andrews 1974), 
is in the circum-nuclear starburst ring. Its reported optical 
coordinates\footnote{http://nedwww.ipac.caltech.edu/} are 
just $\approx 1$\arcsec\ away from the {\it Chandra} position 
of source No.~56. 
We are unable to ascertain whether or not the X-ray source is indeed 
the remnant of SN 1968L. In any case, 
SNe occur at a rate $\sim 1$ per century in the nuclear region. 
With its soft spectrum, inconsistent with 
powerlaw models typical of accreting binaries, source No.~56 
is almost certainly an X-ray SNR.


Two more sources were suggested as X-ray SNR candidates 
by Immler et al.\ (1999), based on VLA radio and {\em ROSAT} HRI data.
In our Table A.1 catalogue, they are sources No.~73 and 86, 
corresponding to H23 and H26 in Immler et al.\ (1999). 
Source No.~73 is located in the bar, 
  in front of the dust lane, $\approx 30$\arcsec\ north-east of the nucleus, 
  and it coincides with the position of a compact radio source 
  (Cowan \& Branch 1985; Cowan et al.~1984). 
Source No.~86 is located near the end of the north-eastern arm. 
Both sources are situated in giant H\,{\footnotesize II} regions.    

The spectra of both sources can be fitted by single-temperature vmekal models. 
For source No.~73, the fit parameters are $kT_{\rm vm} = (4.2\pm0.7)$~keV, 
  $n_{\rm H} = (4.8\pm2.0) \times 10^{20}$~cm$^{-2}$ 
  ($\chi^2_\nu = 0.97$, 42~d.o.f.). 
For source No.~86, we obtain 
  $kT_{\rm vm} = (1.8\pm0.1)$~keV, 
  $n_{\rm H} = (24\pm2) \times 10^{20}$~cm$^{-2}$ 
  ($\chi^2_\nu = 0.60$, 105~d.o.f.).   
In both cases, the fitted temperatures are higher 
than for most SNRs.
More importantly, their spectra are smooth, without any significant 
emission-line features. They could also be well fitted   
  by powerlaw or Comptonised blackbody models, 
  with parameters similar to those typical of XRBs 
in a high-soft state (Table B.4).  
The nature of the two sources is therefore still open to investigation.
We argue that they are more likely to be XRBs 
  rather than young SNRs.  
If so, an estimated emitted luminosity 
$\simgt 3 \times 10^{38}$~erg~s$^{-1}$ 
in the $0.3$--$8.0$~keV band makes source No.~86 
a likely BH candidate.

Among the fainter X-ray sources, 
  No.~84 ($L_{\rm x} \approx 1.5 \times 10^{37}$~erg~s$^{-1}$ 
  in the $0.3$--$8.0$ keV band) 
  coincides with a flat-spectrum 
  radio source (spectral index $0.21\pm0.14$; Cowan et al.\ 1994). 
The radio flux was $(0.39\pm 0.05)$ mJy at 20~cm, and $(0.50\pm0.05)$ mJy at 6~cm  
  in 1990--1992. 
The radio brightness at 6~cm faded by $(33\pm7)$\% between 1983--1984 and 1990--1992. 
The source is probably an evolving radio SNR (Cowan et al.\ 1994);  
the {\it Chandra} detection suggests that source No.~84 may be its X-ray counterpart. 
Unfortunately the X-ray source is not bright enough for detailed spectral modeling. 
Another faint source, No.~100 
  ($L_{\rm x} \approx 6 \times 10^{36}$~erg~s$^{-1}$ in the $0.3$--$8.0$ keV band),  
  coincides with the non-thermal radio source No.~7 in Cowan et al.\ (1994). 
The radio source faded by $\approx (25\pm11)$\% at 6~cm 
  between 1983--1984 and 1990--1992. 

\subsection{X-ray binaries}  
\label{xrb}   

The remaining sources for which we could do meaningful 
spectral modelling do not show any spectral features 
resembling line emission 
 from optically-thin thermal plasma, 
  nor do they have the thermal blackbody-like soft spectra 
  typical of supersoft sources. 
Most of them are likely to be XRBs containing an accreting 
  BH or NS. 
An absorbed powerlaw provided a good fit 
for the X-ray spectra of most sources (Table B.4).
From the best-fit photon indices,  
  we notice that the sources can be divided into two groups. 
The first group (No.~5, 31, 33, 44, 60, 78, 85 and 121)  
  contains sources with hard spectra: all of them have photon indices 
  $\Gamma \approx 1.5$, except for source No.~78, which is even 
harder, with $\Gamma \approx 1.0$.  
Sources in the second group (No.~64, 72, 88, 104 and 113, plus 
sources No.~73 and 86 described in Sect.~3.3)  
  have much steeper (softer) spectra, with $\Gamma > 2$. 
Therefore, we also fitted the spectra of this second group of sources 
with the diskbb and bmc models in {\footnotesize XSPEC}: 
both provided equally acceptable or better fits than a powerlaw model.  
Typical fit temperatures inferred from the diskbb model are 
within the $0.7$--$1.5$~keV range.  
For the bmc model, we obtain fit temperatures $\sim 0.2$--$0.8$~keV 
for the seed photon component.           
   
The luminosity and spectral parameters 
  obtained from the powerlaw, diskbb and bmc models    
  are similar to those found in Galactic XRBs, 
  supporting our interpretation that  
  most of these bright sources are accreting NSs and BHs. 
Galactic XRBs often have two distinguishable X-ray spectral states: 
  the soft state, when the spectrum is  
  dominated by a thermal (blackbody or disk-blackbody) component 
  at a temperature $kT \sim 1$~keV, 
  sometimes with an additional powerlaw component of index $\Gamma \sim 2.5$; 
  and the hard state, characterized by a simple powerlaw spectrum 
  with a photon index $\Gamma \sim 1.5$.
The bright sources in M\,83 can also be classified into two states 
  with analogous spectral properties: roughly half of them appear 
to be in the soft state and half in the hard state.  
A photon index $\Gamma \approx 1.0$ inferred for source No.~78 
is unusually hard but not unique among XRBs: for example, 
the transient high-mass XRB SMC X-2 also 
  has $\Gamma \approx 1.0$ and luminosity $\approx 10^{38}$ erg s$^{-1}$ 
  in its high state (Corbet et al.\ 2001). 


\subsection{A background radio galaxy}  
\label{fr2}   

Source No.~39 is certainly identified as a background object 
  because it is coincident with an inverted-spectrum radio source 
  (source 3 in Cowan \& Branch 1985; Cowan et al.~1994), 
  which is thought to be the core of an FR II radio galaxy. 
(This identification was also noted by Stockdale et al.\ 2001).
In addition to the flat-spectrum radio core, 
  VLA 6 cm and 20 cm images also show two distinct radio lobes, 
  at $\approx 25$\arcsec\ north-west and south-east of the core 
  (Cowan et al.\ 1994), 
The lobes are brighter than the core,  
  and have a steep spectrum (spectral indices $\approx -1$).
 
From the {\it Chandra} data, 
  we estimate an emitted X-ray flux 
  $f_{\rm x} = 2.2 \times 10^{-14}$ erg cm$^{-2}$ s$^{-1}$ 
  in the 0.5--3.0 keV band, and 
  $f_{\rm x} = 5.0 \times 10^{-14}$ erg cm$^{-2}$ s$^{-1}$ 
  in the 0.3--8.0 keV band. 
This corresponds to a monochromatic luminosity at 2 keV 
  (defined like in Fabbiano et al.\ 1984) 
  $l_{\rm x} = 5.5 \times 10^{24}$ 
  $(d_{\rm L}/10^{9} {\rm pc})^2$ erg s$^{-1}$ Hz$^{-1}$, 
  where $d_{\rm L}$ is the luminosity distance. 
The monochromatic luminosity of the radio nucleus at 5GHz is (Cowan et al.\ 1985) 
  $l_{\rm RN} = 1.6 \times 10^{30}$ 
  $(d_{\rm L}/10^{9} {\rm pc})^2$ erg s$^{-1}$~Hz$^{-1}$. 
These two values satisfy the linear correlation 
  for FRII galaxies and quasars 
  in the (log $l_{\rm x}$, log $l_{\rm RN}$) plane 
  (Fig.~9 in Fabbiano et al.\ 1984), 
  regardless of the assumed distance.

We can estimate the distance from the fact  
  that the radio luminosity of an FRII radio galaxy is always 
  $L_{\rm 20cm} \simgt 10^{32}$ erg s$^{-1}$ Hz$^{-1}$, 
  and $L_{\rm 6cm} \simgt 10^{32}$ erg s$^{-1}$ Hz$^{-1}$, 
  for a Hubble constant $H_0 = 75$ km s$^{-1}$ Mpc$^{-1}$ 
  (eg, Owen \& Ledlow 1994; Kembhavi \& Narlikar 1999).
If the radio galaxy seen behind M\,83 is  
  at redshift $z = 1$ 
  (corresponding to $d_{\rm L} = 4.7$ Gpc for $q_0 = 0.5$ 
   and $H_0 = 75$ km s$^{-1}$ Mpc$^{-1}$), 
   its monochromatic radio luminosities 
   are $L_{\rm 20cm} = 1.5 \times 10^{32}$ erg s$^{-1}$ Hz$^{-1}$,
   and $L_{\rm 6cm} = 0.7 \times 10^{32}$ erg s$^{-1}$ Hz$^{-1}$ (Cowan et al.\ 1994). 
Its emitted X-ray luminosity is then $L_{\rm x} = 1.3 \times 10^{44}$ erg s$^{-1}$ 
   in the $0.3$--$8.0$ keV band. 
Values of $L_{\rm x} \sim 10^{44}$ erg s$^{-1}$ are typical of FRII radio galaxies. 
We conclude that the source is located at a redshift $z \simgt 1$. 
(We have no elements to determine an upper limit to its distance.)

The angular diameter distance is only a weak function of redshift, 
  for $z \simgt 0.75$: the inferred total size of the radio lobes 
projected in the plane of the sky 
  is $\approx 250$--$280$ kpc. 
This is an average value for this class of objects (Blundell et al.~1999). 


\section{Variability of the sources}  

We searched for variabilities of the brightest sources 
  on timescales between 10 and 50,000~s. 
Shorter periods cannot be investigated, 
  because the ACIS observation was carried out in ``Timed Exposure'' mode, 
  collecting photons and reading them out every $\approx 3.2$~s. 
We were aware that spurious periods 
  could be found at $\approx 707$~s and $\approx 1000$~s 
  due to dithering.
After extracting lightcurves with standard CIAO routines,  
  we carried out discrete Fourier transforms  
  to obtain the power density spectra. 
Null results were obtained for most sources;   
  however, periods of 174.9~s and 201.5~s were found 
  for sources No.~33 and 113 respectively.  
We checked 
  that neither the background emission around those sources 
  nor other bright nearby sources showed similar periodicities. 
Hence, we conclude that the periodic 
variabilities are probably intrinsic to the X-ray sources.

\begin{figure}
\begin{center} 
\vspace*{0.3cm} 
\epsfig{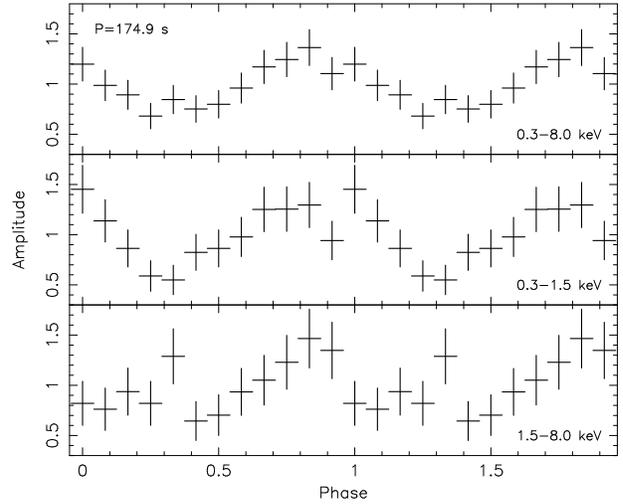}  
\end{center}
\caption{ 
Lightcurves of source No.~33 
  in the total, soft and hard bands, 
  normalized to the mean count rates, 
  folded on a period of 174.9~s. }
\label{fig:light_curve33}
\end{figure}

\begin{figure}
\begin{center} 
\vspace*{0.3cm} 
\epsfig{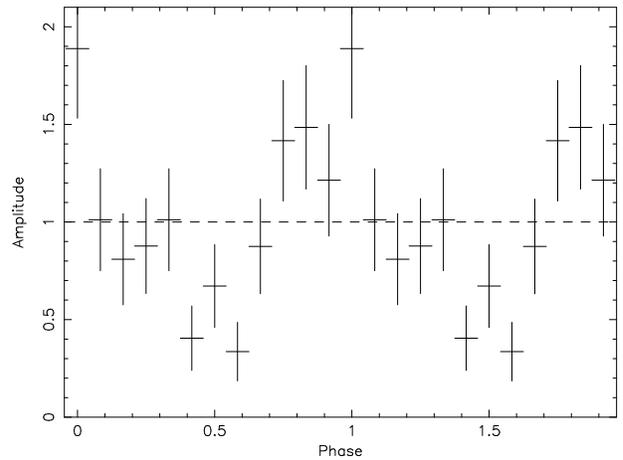}  
\end{center}
\caption{ 
  Lightcurve of source No.~113, 
    normalized to the mean count rate, 
  folded on a period of 201.5~s. }
\label{fig:light_curve113}
\end{figure}

We folded the two X-ray lightcurves according to those periods 
  and found large amplitude variations 
  (Fig.~\ref{fig:light_curve33} and \ref{fig:light_curve113}). 
Their pulse amplitudes are approximately 0.5 of the mean levels, 
  similar to those seen in Galactic X-ray pulsars. 
The spectra of the two sources are well fitted 
  by a simple powerlaw (Table B.4), 
  consistent with the spectra of typical X-ray pulsars. 
These two systems may be Be XRBs 
  (ie, accreting NSs in eccentric orbit around Be stars) in outburst,  
  similar to those often observed in the Magellanic Clouds 
  (eg, Negueruela 1998; Laycock et al.\ 2002).
If so, repeated observations should show luminosity variations 
  over the orbital period, 
  which in Galactic Be XRBs is typically $\sim 100$~d 
  for spin periods $\sim 100$~s (Corbet 1986).

                          
\section{Statistical properties of the source population}

\subsection{Brightness distribution}  

\begin{figure} 
\begin{center}  
\vspace*{0.5cm}
\psfig{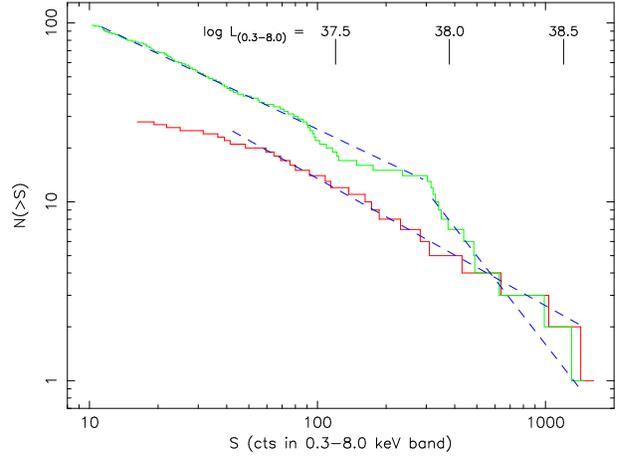}  
\end{center}
\caption{
Red histogram: cumulative luminosity distribution of the discrete sources 
    in the inner region ($d < 60\arcsec$~from the galactic center), 
together with a powerlaw approximation (index $-0.7$). 
Green histogram: cumulative luminosity distribution 
of the sources in the outer disk, together with a broken 
powerlaw approximation (break at an emitted luminosity 
of $8 \times 10^{37}$~erg~s$^{-1}$ for the assumed spectral model; 
indices of $-1.6$ and $-0.6$ above and below the break).
 }
\label{fig:lumd}
\end{figure}

Among the 127 sources in the S3 chip, 
  28 are located in the inner disk, defined as 
the starburst nuclear region plus the bar, 
within 60\arcsec~of the X-ray center;  
  17 of them are within 16\arcsec~from the galactic center. 
The luminosity functions 
  of the 28 ``inner disk'' sources and of the 99 ``outer disk'' sources  
  differ substantially (see also Soria \& Wu 2002). 
Above the completeness limit ($\approx 50$ counts for the nuclear region), 
  the inner-disk sources have a simple powerlaw luminosity distribution 
  with an index of $-0.7$ (Fig.~\ref{fig:lumd}).  
In contrast, the luminosity distribution of the other sources 
shows a deficiency of bright objects above 300 counts. 
For a distance of 3.7~Mpc and an assumed absorbed powerlaw spectral model 
  ($\Gamma = 1.7$, $n_{\rm H} = 10^{21}$~cm$^{-2}$), 
  300 counts correspond to an emitted luminosity of $8 \times 10^{37}$~erg~s$^{-1}$.  
If we model the luminosity function with a broken powerlaw 
  with a break at about 300 counts,   
  we obtain powerlaw indices of $-0.6$ and $-1.6$ above and below the break. 
There may be another break at about 100 counts, 
  but a more quantitative analysis is impossible  
  because of the small number of sources.
The detection limit is $\approx 12$ counts, however we estimate 
that the sample of outer-disk sources is complete only down to 
$\approx 20$ counts: a few soft sources with $\simlt 20$ counts 
may be missed if located in star-forming regions along the spiral arms 
(characterised by diffuse soft emission).

The luminosity distribution of bright X-ray sources 
  reflects the star-formation activity of the host galaxy  
  and/or the host environments in the recent past 
  (Wu 2001; Wu et al.\ 2003; Prestwich 2001; Kilgard et al.\ 2002).   
For example, there is a lack of
  bright sources in relatively quiescent galaxies 
  (eg, M\,31, Kong et al.\ 2002) 
  while they are much more abundant in starburst (eg, M~82, Matsumoto et al.\ 2001) 
  and interacting galaxies (eg, NGC~4038/4039, Zezas et al.\ 2002). 
Such differences can also be observed between different components of the same 
galaxy: for example, in M\,81, the sources in the bulge 
  show a broken-powerlaw luminosity function typical of quiescent galaxies,  
  but the sources in the disk show a simple powerlaw luminosity function 
  similar to those of starburst and interacting galaxies  
  (Tennant et al.\ 2001; Swartz et al.\ 2003).  

We may explain the flat powerlaw luminosity function 
of the nuclear sources in M\,83
  in terms of current starburst activity. 
The explanation of the luminosity function of the outer disk sources   
  is less straightforward. 
Some of the sources might have been formed in the spiral arms  
  as a result of continuous, density-wave induced star formation; 
  others might be remnants from previous starburst epochs, 
  perhaps due to past close encounters of M\,83 with its neighbors.
We speculate that this may explain the break-like features 
seen in the luminosity function between $\approx 3 \times 10^{37}$ 
and $10^{38}$ erg s$^{-1}$. We also note that 
  the faint end of the luminosity functions for inner and outer disk sources 
  have similar slopes. If we interpret the faint end  
  as the quasi-stationary part of the source distribution,   
  less prone to transient perturbations such as starburst episodes,  
  then their similarity suggests 
  that the underlying processes determining the luminosity functions  
  in the inner and outer disk are the same. 

Whether or not some of the bright sources are transients 
does not affect 
   the shape of the luminosity function, provided 
   that the transitional probability between the high and the low 
   states is quasi-stationary. In other words, the transient 
   nature of the sources will not introduce features in a luminosity 
   function unless the transitional timescale is comparable to the 
   evolutionary timescale of the luminosity function. Many 
   XRBs in our Galaxy show spectral transitions on a few months/years 
timescale, while the luminosity function should evolve on a much longer 
   timescale. 

 
\subsection{Spectral and color distribution}  

We constructed color-color diagrams to demonstrate  
  that the discrete sources can be classified into 
physically distinct groups.  
We considered two different definitions of the color indices: 
  the conventional indices [$(M-S)/(M+S)$,\,$(H-S)/(H+S)$], 
  used for example by Swartz et al.\ (2002) 
  in the study of M\,81 sources,  
  and the alternative color indices [$(H-M)/(H+M+S)$,\,$(M-S)/(H+M+S)$] 
  suggested by Prestwich et al.\ (2003).  

\begin{figure}[t]
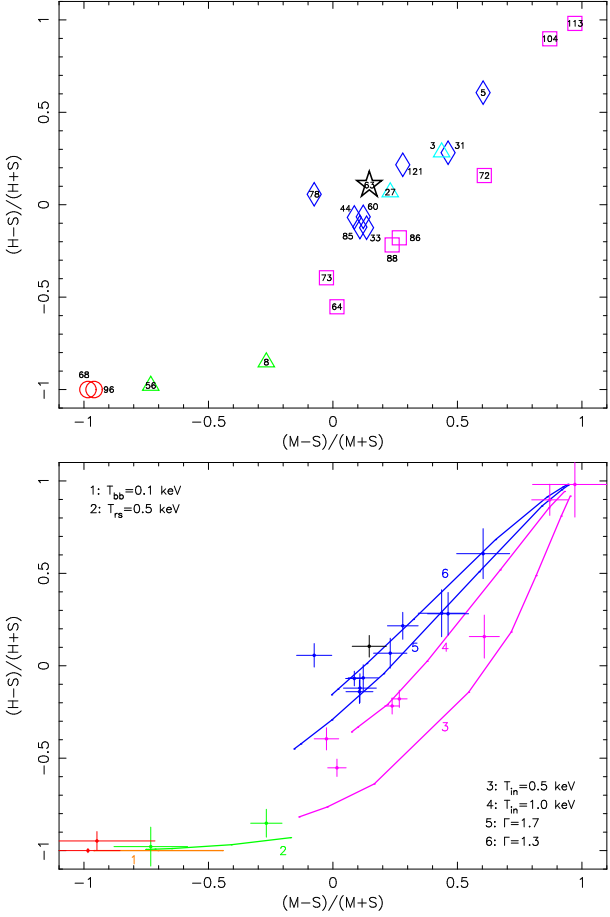

\begin{center} 
\vspace*{0.35cm} 
\epsfig{figure=m83_color23_dg1b.ps,width=6.0cm,angle=270} \\
\vspace{0.1cm} 
\epsfig{figure=m83_color23_dg2b.ps,width=6.0cm,angle=270} \\
\end{center}
\caption{Top panel: 
  Color-color diagram of the 22 sources in M\,83 
  (the source associated with the background radio galaxy is not included)
  for which individual spectral fitting is available.   
Sources with a featureless, hard powerlaw spectrum are plotted as blue diamonds. 
Sources with emission lines on top of a hard powerlaw spectrum 
  are plotted as cyan triangles. 
Magenta squares are sources with a featureless soft spectrum. 
Green triangles are sources characterised by soft, optically-thin thermal 
plasma emission. 
Red circles are supersoft sources (blackbody-like emission at temperatures $< 80$ eV). 
The black star is the galactic nucleus. 
The numbers refer to the source catalogue in Table A.1.
Bottom panel: the colors of those same 22 sources 
(plotted here with their error bars) are compared with those 
expected from some simple spectral models: 
  blackbody, disk-blackbody, powerlaw, and optically-thin thermal plasma. 
The column density increases along each model curve 
from bottom left to top right, 
  from $n_{\rm H} = 4 \times 10^{20}$~cm$^{-2}$ 
  (line-of-sight foreground absorption for M\,83) 
  to $n_{\rm H} = 7.5 \times 10^{21}$~cm$^{-2}$ 
  for the blackbody and optically-thin thermal plasma models, 
  and to $n_{\rm H} =  2 \times 10^{22}$~cm$^{-2}$ 
  for the powerlaw and disk-blackbody models.
}
\label{fig:col_col_dg1}
\end{figure}

\begin{figure}[h]
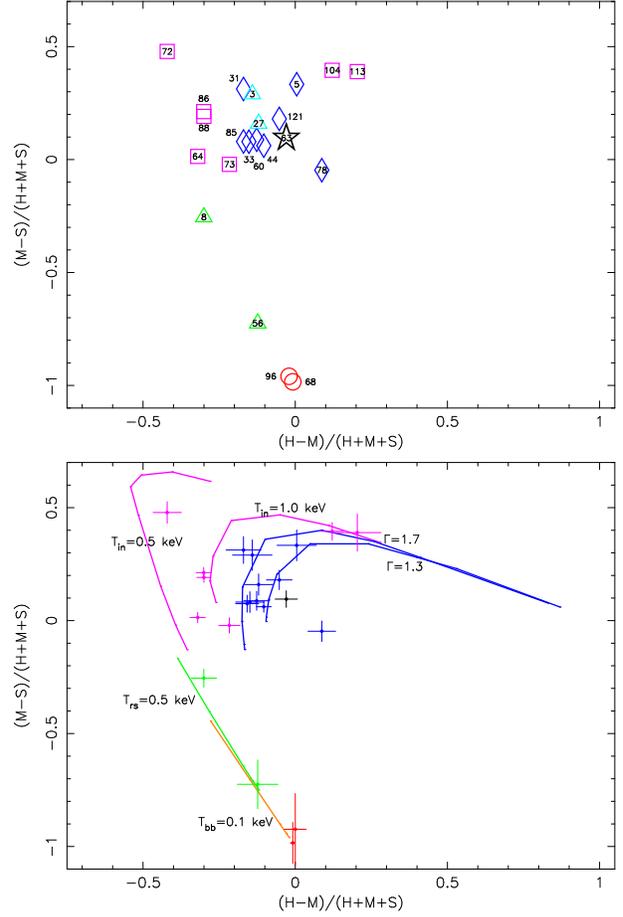

\begin{center} 
\vspace*{0.35cm} 
\epsfig{figure=m83_color23_ap1b.ps,width=6.0cm,angle=270} \\
\vspace{0.1cm}
\epsfig{figure=m83_color23_ap2b.ps, width=6.0cm,angle=270} \\
\end{center}
\caption{
  Color-color diagrams for the alternative color indices.  
  The symbols and colors of the 22 sources, 
    and the superimposed spectral models 
    are the same as in Fig.~\ref{fig:col_col_dg1}. 
  For each spectral model, 
    the absorbing column density increases 
    from the bottom to the top of the representative curves.
}
\label{fig:col_col_ap1}
\end{figure}

We first considered the sources 
  for which we have obtained individual spectral fits (Sect.~3), 
  as indicators of the physical nature of the sources 
  in the various regions of the color-color diagrams   
(top panels of Fig.~\ref{fig:col_col_dg1} and \ref{fig:col_col_ap1}).
We then compared their colors  
with those expected from a sample of fundamental spectral models 
(bottom panels of Fig.~\ref{fig:col_col_dg1} and \ref{fig:col_col_ap1}). 
The spectral models that we considered are: 
  powerlaws with photon indices 
  $\Gamma = 1.3$ and $\Gamma = 1.7$ 
  (characteristic of XRBs in the hard state); 
  disk-blackbodies with $kT_{\rm in} = 0.5$ keV and $kT_{\rm in} = 1.0$ keV 
  (typical of XRBs in the soft state); 
  blackbody with $kT_{\rm bb} = 0.1$ keV 
  (supersoft sources);
  optically-thin, single-temperature thermal plasma at $kT_{\rm rs} = 0.5$ keV 
  (typical of SNRs).  
Along each model curve, the column density increases from the bottom 
to the top, 
  from $n_{\rm H} = 4 \times 10^{20}$~cm$^{-2}$ 
  (line-of-sight foreground absorption for M\,83) 
  to $n_{\rm H} =  7.5 \times 10^{21}$~cm$^{-2}$ 
  for the blackbody and optically-thin thermal plasma models, 
  and to $n_{\rm H} =  2 \times 10^{22}$~cm$^{-2}$ 
  for the powerlaw and disk-blackbody models. 
 
                         
\begin{figure}
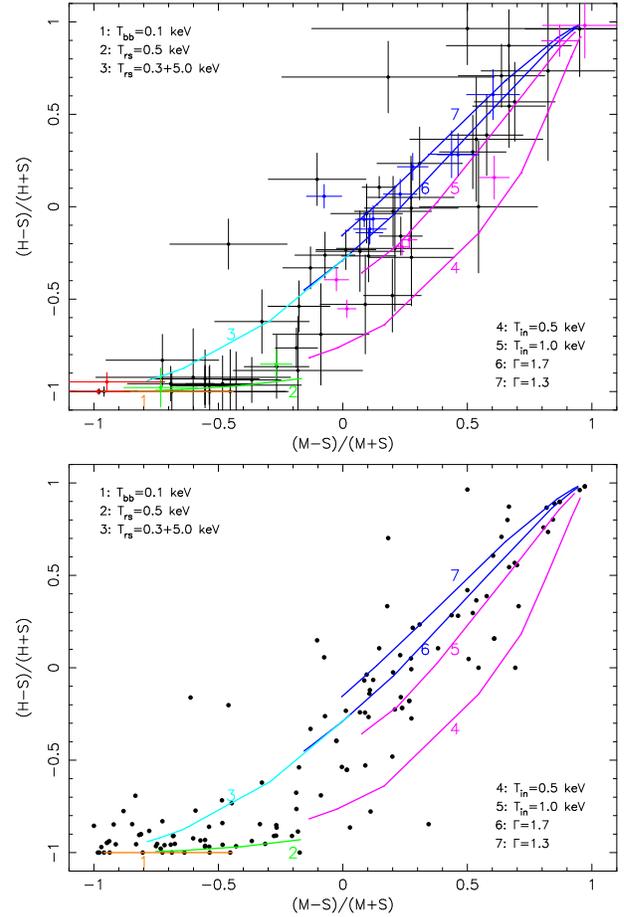

\begin{center} 
\vspace*{0.35cm} 
\epsfig{figure=m83_colplot3b.ps, width=6.0cm,angle=270} \\
\vspace{0.1cm} 
\epsfig{figure=m83_colplot4b.ps, width=6.0cm,angle=270} \\
\end{center}
\caption{Top panel: 
  color-color diagram for sources brighter than $\approx 10^{37}$ erg s$^{-1}$,  
  with respective error bars. 
Bottom panel: color-color diagram for the all sources in the S3 chip, 
  without error bars.
In addition to the models shown in Fig.~\ref{fig:col_col_dg1}, 
  we have also overplotted a two-temperature thermal plasma model 
  ($kT_{\rm rs,1} = 0.3$ keV, $kT_{\rm rs,2} = 5.0$ keV), 
  normalised so that 
  the high-temperature component accounts for 25\% of the emitted flux. }
\label{fig:col_col_dg2}
\end{figure}

\begin{figure}
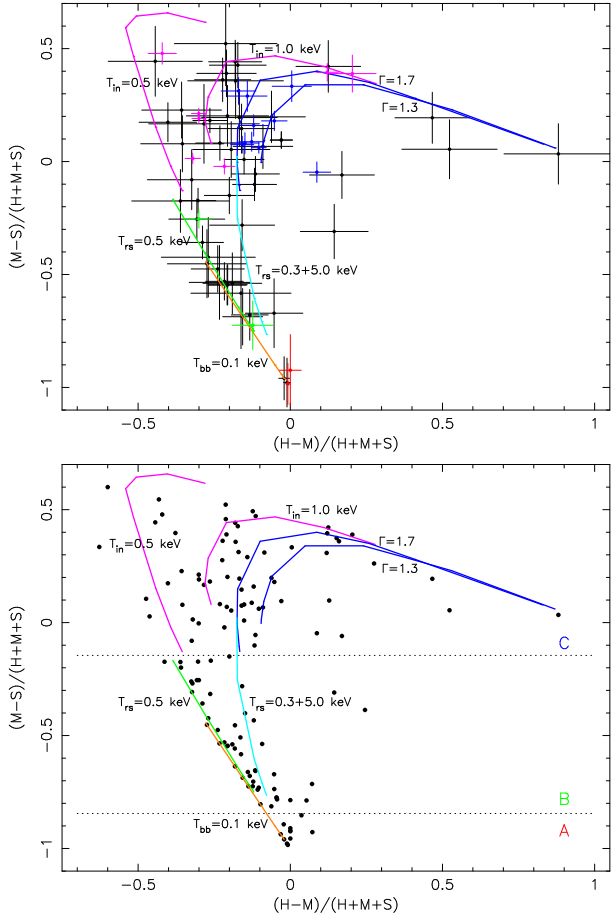

\begin{center} 
\vspace*{0.35cm} 
\epsfig{figure=m83_colplot1d.ps,width=6.0cm,angle=270} \\
\vspace{0.1cm}
\epsfig{figure=m83_allcolors2.ps, width=6.0cm,angle=270} \\
\end{center}
\caption{ 
 Same as Fig.~\ref{fig:col_col_dg2} for the alternative color indices. 
 The three groups (A, B, C) are separated by the horizontal dotted lines.   
} 
\label{fig:col_col_ap2}
\end{figure}

Different types of sources are clearly distinguished 
in both color-color diagrams. 
With the conventional choice of color indices 
  (Fig.~\ref{fig:col_col_dg1}) one 
  can separate XRBs (with $(M-S)/(M+S) \simgt -0.2$) 
from the other sources. For the XRBs, it is then possible 
to distinguish whether a source is intrinsically hard, 
or simply more absorbed. 
However, the alternative color indices 
  (Fig.~\ref{fig:col_col_ap1}) 
  appear to be better in separating supersoft sources, SNRs and XRBs.
When we plot the colors indices of all the sources found in the S3 chip 
(Fig.~\ref{fig:col_col_dg2} and \ref{fig:col_col_ap2}), 
  the advantage of the alternative choice of indices is more obvious,  
  especially for the faint sources.

We may roughly divide the alternative color-color diagram 
  into three regions, each dominated by a different source type.
Here and hereafter we refer to the sources in those three regions 
as group A, B and C respectively (Fig.~\ref{fig:col_col_ap2}).
Group A is dominated the ``classical'' supersoft sources,  
  which have negligible emission in the medium and hard X-ray 
bands and therefore cluster around (0,\,$-1$) regardless of absorption\footnote{
  For an operative way of identifying and classifying soft and supersoft sources 
    in {\it Chandra} ACIS observations, see Di\,Stefano \& Kong (2003).}.
Group B includes all the sources located at $-0.85 \simlt (M-S)/(S+M+H) \simlt -0.15$. 
Most of them are found along a straight line in the diagram, 
with only a narrow spread in $(H-M)/(S+M+H)$. This is consistent with emission from 
single-temperature optically-thin thermal plasma at $kT_{\rm rs} \approx 0.4$--$0.8$ keV, 
typical of SNRs (Prestwich et al.\ 2003).
Only a few group B sources have slightly harder spectra, 
consistent with emission from a two-temperature optically-thin plasma   
  with $kT_{\rm rs,1} \approx 0.5$ keV and $kT_{\rm rs,2} \approx 5$--$10$ keV. 
Alternatively, but less likely, they may have a composite spectrum  
  consisting of a very absorbed, hard powerlaw and a (dominant) reprocessed 
  soft component, as observed in some Galactic NSs (see Sect.~5.4).  

Group C sources are located at $-0.15 \simlt (M-S)/(S+M+H) \simlt 0.6$, 
(Fig.~\ref{fig:col_col_ap2}) 
with a much broader spread in $(H-M)/(S+M+H)$. This corresponds 
to a broad spread in both their absorbing column densities 
and their intrinsic spectral hardness.
We believe that most of them are XRBs, although a few 
hard, highly-absorbed sources may be background AGN.
All sources with a disk-blackbody, disk-blackbody plus powerlaw, 
or simple powerlaw spectra fall in this region of the diagram. 

Incidentally, we notice that the emission-line sources No.~3 and 27 
(discussed in Sect.~3.3) have colors consistent with 
group C sources (Fig.~\ref{fig:col_col_ap1} and~\ref{fig:col_col_ap2}): this 
is in agreement with our argument that they are accreting 
systems surrounded by a photo-ionized nebula or stellar wind, 
rather than SNRs.  
We also note that only 5 sources are clearly outliers, 
with much harder spectra than those of typical XRBs or SNRs. 
They could be analogous to the bright ($L_{\rm x} \sim 10^{38}$ erg s$^{-1}$) 
high-mass XRBs SMC X-2 (Corbet et al.\ 2001) and LMC X-4 (La Barbera et al.\ 2001), 
whose X-ray spectra have powerlaw components with $\Gamma \simlt 1$. 
 
XRBs in a soft and hard state are well separated 
in this kind of color-color diagrams. The number 
of bright sources detected in each state is comparable, 
suggesting that the bright XRBs in this galaxy are not 
preferentially found in either spectral state.
It is less easy to separate BH from NS XRBs, 
or high-mass from low-mass XRBs, based on colors alone. 
As a comparison with bright sources in the Galaxy and Magellanic Clouds, 
  the high-mass XRBs LMC X-1 and LMC X-3 are almost always detected 
  in a soft spectral state; 
  on the other hand, Cyg X-1 is usually found in a hard state 
  (with occasional transitions to a soft state), 
  and most transient Be XRBs have a hard spectrum in outburst. 
  BH XRBs with low-mass companions (eg, GX339$-$4, XTE J1550$-$564) 
  and NS low-mass XRBs (eg, 4U 1728$-$34) 
  also show transitions between soft and hard states.
(See eg, Lewin et al.~(1995) for a review).
However, the spectral 
and color differences between the two ``canonical'' 
states are generally larger in BH XRBs than in NS XRBs (eg, Sunyaev 2001; 
Sunyaev \& Revnivtsev 2000): BHs are harder when in the hard state, 
and softer when in the soft state. 
Hence, BH XRBs should have larger color excursions 
within the group C sources. Repeated observations 
of the source colors over timescales of a few months/years 
may help distinguish NS from BH systems. 
Instead, we do not expect to see 
color transitions between different groups.


 
                          
\subsection{Correlation with H$\alpha$ emission}

\begin{figure} 
\begin{center}  
\vspace*{0.35cm}
\psfig{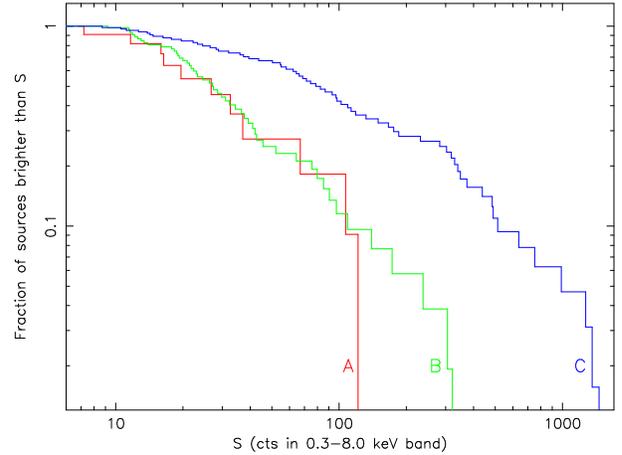}  
\end{center}
\caption{Cumulative luminosity distributions 
of the three groups of sources (normalized to their total numbers) 
identified in Sect.~5.3. 
The distributions for groups A, B and C are plotted in red, green 
and blue respectively. 
 }
\label{fig:lumfun3}
\end{figure}

The cumulative luminosity distributions of group A, B and C sources 
are not identical (Fig.~\ref{fig:lumfun3}).   
Group A (supersoft) and B (soft) sources   
  have a much lower count rate: only $\approx 10$\% of them have 
  $\simgt 100$ cts ($\approx 0.002$ cts s$^{-1}$, 
corresponding to $\approx 3 \times 10^{37}$ erg s$^{-1}$). 
The luminosity function of the group B sources is 
approximately a powerlaw with index $-1$.
Most of the bright sources belong to group C: $\approx 40$\% 
of the sources in this group have X-ray luminosities 
  $\simgt 3 \times 10^{37}$ erg s$^{-1}$.  
This fraction increases to $\approx 50$\% 
  when we subtract the estimated background AGN contribution.

\begin{figure} 
\begin{center}  
\vspace*{0.35cm}
\psfig{figure=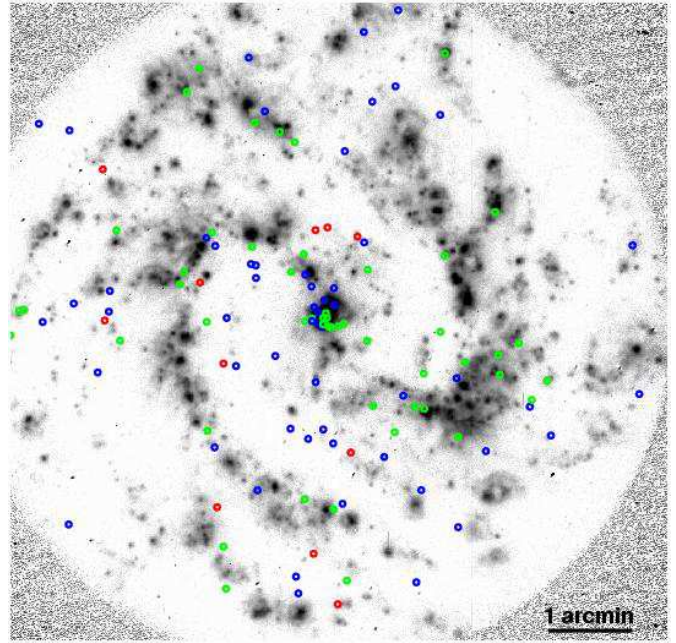,width=8.8cm}  
\end{center}
\caption{
  Location of the discrete X-ray sources classified according 
to their color group (red for group A sources; green for group B; 
blue for group C), overplotted  
     onto a continuum-subtracted H$\alpha$ image of M\,83.  
  North is up, East is left.
  The H$\alpha$ image was taken by Stuart Ryder 
     from the Anglo-Australian Telescope.  
  }
\label{fig:ha_map}
\end{figure}

\begin{figure} 
\begin{center}  
\vspace*{0.35cm}
\psfig{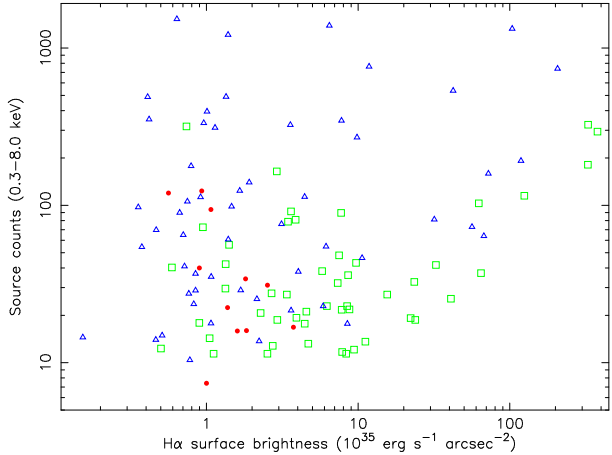}  
\end{center}
\caption{
X-ray source count rate versus  
the H$\alpha$ surface brightness in the region 
around the source (average value in a circle of radius $5\arcsec$).  
Here red circles = group A, green squares = group B, 
blue triangles = group C sources.
The lack of detected sources near the bottom right corner of the diagram 
is due to the completeness limit of our X-ray sample 
in regions of high H$\alpha$ emission: those regions are also 
characterized by strong, diffuse X-ray emission, so that faint X-ray 
sources cannot be detected. 
}
\label{fig:xray_ha}
\end{figure}

We investigated the spatial correlation between different groups of 
  X-ray sources and the H\,{\footnotesize II} regions, 
  which define the spiral arm structure and are 
indicators of young stellar populations.  
Overplotting the X-ray sources on a continuum-subtracted H$\alpha$ map of M\,83   
  (Fig.~\ref{fig:ha_map}) 
  shows that group B sources (green circles) 
  are more closely associated with H$\alpha$ emission 
  (mostly found in the starburst nucleus and the spiral arms), 
  than group A and C sources. 
We quantified this association 
  by calculating the average H$\alpha$ surface brightness  
  within a circle of 5\arcsec~radius 
  around the position of each X-ray source,
  and plotting it as a function of the X-ray flux, 
  for each group of sources (Fig.~\ref{fig:xray_ha}).
The data do not show any correlation 
  between X-ray count rates and H$\alpha$ surface brightness for any of the three groups. 
The lack of faint X-ray sources associated with very bright H$\alpha$ emission 
  (near the bottom right of the plot) 
  is due to the incompleteness of the X-ray sample: 
  bright H$\alpha$ emission is usually associated with strong diffuse X-ray emission  
  (eg, in the starburst nucleus), where faint X-ray sources are more easily unresolved. 

\begin{figure} 
\begin{center}  
\vspace*{0.35cm}
\psfig{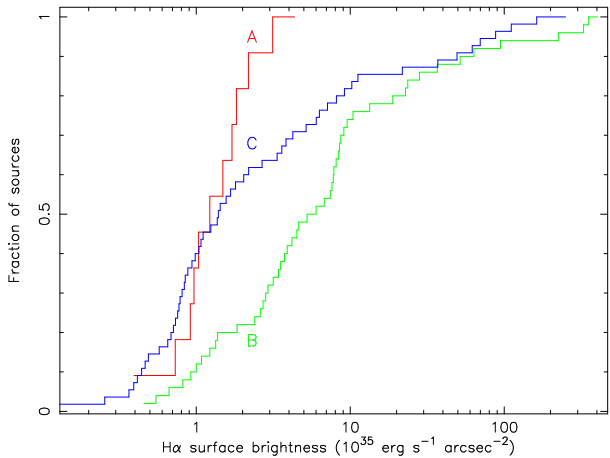}  
\end{center}
\caption{Cumulative distributions of the H$\alpha$ surface brightness 
associated to each group of X-ray sources. 
 }
\label{fig:ha_distr}
\end{figure}

We then examined (Fig.~\ref{fig:ha_distr}) the cumulative distribution 
of the H$\alpha$ surface brightness associated with the three groups 
of X-ray sources (the total number of sources is normalized to 1 
for each group). This shows a significantly stronger H$\alpha$ emission 
associated with group B sources: almost 80\% of them 
are located in regions with H$\alpha$ surface brightness 
$> 2 \times 10^{35}$ erg s$^{-1}$ arcsec$^{-2}$, 
compared to only 40\% of group C sources.
This is further evidence that group B sources come from 
a younger population. This difference is particularly 
evident in regions of intermediate H$\alpha$ brightness, 
typical of H\,{\footnotesize II} regions in the spiral arms.
For regions of very bright H$\alpha$ emission 
($\simgt 2 \times 10^{36}$ erg s$^{-1}$ arcsec$^{-2}$) 
B and C sources have similar behavior: this corresponds 
to the starburst nucleus, where we expect to find  
a high concentration of both SNRs and high-mass XRBs.

Sources of group A (supersoft) 
are found only in regions of low H$\alpha$ emission: 
this is consistent with an old population. 
However, it could also be due to a selection effect: 
  supersoft sources embedded in a spiral arm or in the starburst nucleus 
  (and hence associated with bright H\,{\footnotesize II} regions) 
  are heavily absorbed and may not detected easily in the soft X-ray band. 
The small number of group A sources (11) 
  also makes statistical errors larger.

\subsection{X-ray SNRs and other possible soft sources}

The previous results are consistent with the interpretation 
of the supersoft sources as old 
or intermediate-age nuclear-burning white dwarfs, 
and with the identification of the B sources 
as young SNRs (Prestwich et al.\ 2003).
If this is the case, we expect their luminosity 
to remain constant over many years. Repeated deep 
observations of M\,83 over the next few years 
will verify this conjecture. 

We also looked for spatial correspondences 
between the X-ray sources and the optically-identified 
SNRs (Blair \& Long 2003, in preparation). The latter were identified  
with the [S\,{\footnotesize II}]/H$\alpha$ criterion 
(Blair \& Long 1997), suitable to detect evolved remnants 
of all types over a lifetime of $\approx 20,000$--$30,000$ yr 
(eg, Pannuti 2002 and references therein). 
Seven of the X-ray sources 
classified in group B are located within 5\arcsec~of 
an optical SNR; two 
C sources and one A source are also within 5\arcsec~of 
an SNR. However, most of the optical SNRs are not associated 
with any X-ray sources, and vice versa. 
Little overlapping between X-ray and optically-selected SNRs 
was also noticed in other nearby galaxies (Pannuti et al.\ 2002).
Optical surveys tend to select a larger fraction of SNRs 
in low-density regions (hence, they tend to select 
Type Ia events), while the brighter X-ray SNRs are 
those found in denser star-forming regions.
On the other hand, optical surveys based on the 
[O\,{\footnotesize III}]/H$\alpha$ criterion 
would select mostly remnants from core-collapse 
events (Type II, Ib,c) in young stellar populations. 
Correlations with X-ray sources would provide 
a more constraining test for their age and nature.
This search is beyond the scope of this paper.

\begin{figure}
\begin{center}
\psfig{figure=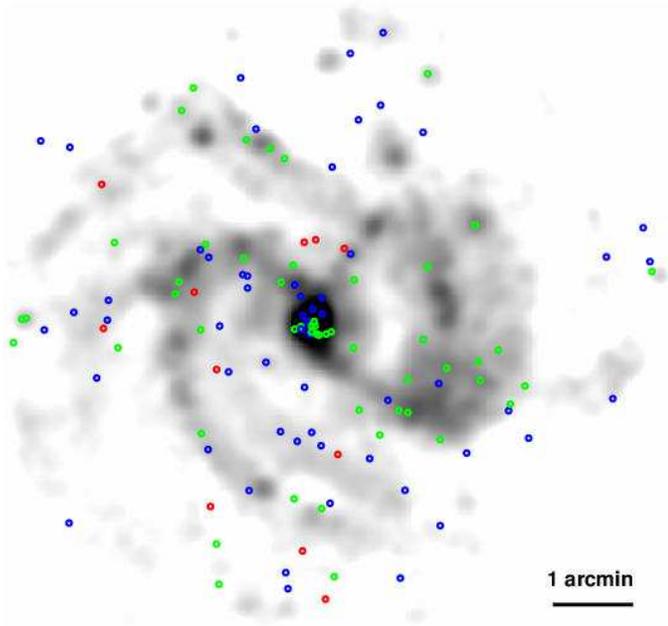,width=9.0cm} 
\end{center}
\caption{Group A, B and C sources (green, red and blue circles respectively) 
overplotted on a map of the radio continuum emission at 6.3cm. 
The radio map is a combination of data 
from the VLA and Effelsberg radio telescopes, at a resolution of 12\arcsec, 
and was produced by S.\ Sukumar, R.\ Allen, R.\ Beck \& N.\ Neininger.}
\label{fig:radiomap}
\end{figure}

A detailed study of spatial correlations between 
X-ray and radio sources is also left to further work. 
A preliminary investigation shows that the continuum 
radio emission at 6.3 cm correlates well with 
the distribution of group B X-ray sources (Fig.~\ref{fig:radiomap}). 
Diffuse thermal radio emission in spiral galaxies is usually 
associated with H\,{\footnotesize II} regions; 
diffuse non-thermal radio emission  
is thought to be produced by cosmic-ray electrons 
accelerated by SNRs. An association between diffuse radio 
emission and group B X-ray sources would strengthen 
their interpretation as SNRs, or at least as 
a young population.

In addition to young SNRs, other classes of sources 
could in principle show X-ray colors consistent 
with our group B classification.
For example, a blackbody or disk-blackbody spectrum 
with $kT \approx 0.1$--$0.2$ keV would  
have the correct X-ray colors. 
Standard accretion-disk models show that a $10$-$M_{\odot}$ BH 
would have to be accreting at a rate $\simlt 10^{-3} \dot{m}_{\rm Edd}$ 
to have a temperature $kT \simlt 0.2$ keV 
at the inner edge of the disk, if it extends 
to the innermost stable orbit of the Schwarzschild geometry.
However, its luminosity would be $\simlt 10^{36}$ erg s$^{-1}$, 
too faint to be detected in our M\,83 observation.
Besides, typical BH candidates in the Galaxy tend to have a harder 
X-ray spectrum when detected at low accretion rates: their 
emission is then dominated by photons Compton-scattered 
in a hot corona, rather than direct emission from 
the accretion disk.
Intermediate-mass BHs ($M \sim 10^3$--$10^4 M_{\odot}$) 
in a high state would have a thermal blackbody-like spectrum 
in the correct temperature range, but their luminosity would be $> 10^{41}$ 
erg s$^{-1}$, not observed in M\,83.

Transient X-ray pulsars with large photospheres 
may be a candidate 
for some of the soft sources. Some systems, such as 
RX J0059.2$-$7138 (Hughes 1994) are known to have 
a very soft ($kT \approx 40$ eV) blackbody component 
in addition to their typical powerlaw component.  
The soft X-ray emission is produced by the Compton 
downscattering of harder X-rays emitted 
from near the surface of the compact object.
With its emitted X-ray luminosity $\sim 10^{37}$ 
erg s$^{-1}$, RX J0059.2$-$7138 would 
be a group B source in our classification scheme.
Time-variability studies are required to distinguish 
this class of transient X-ray pulsars from X-ray SNRs.
Finally, old low-mass neutron star XRBs may have an X-ray 
spectrum dominated by a soft thermal component 
at $kT \approx 100$ eV, caused by the reprocessing 
of hard X-ray photons by the inner edge of an accretion disk. 
An example of this type of sources, which would also 
belong to group B, is the Galactic X-ray pulsar 
Her X-1 (eg, Mavromatakis 1993).

                             

                             
\subsection{Soft sources in M\,83 and M\,81}    

Another clue to determine the nature of the group B sources 
  comes from a comparison of the X-ray colors of the discrete sources 
  detected in other galaxies. 
We considered M\,81, as it is another nearby spiral galaxy  
  but it does not have a starburst nucleus like that of M\,83. 
The foreground absorption in the direction of the two galaxies 
is very similar, hence this does not affect the hardness 
and color comparison. 
In order to compare our results with those of Swartz et al.\ (2002, 2003), 
  we used the {\it Chandra} ACIS count rates for the M\,83 sources 
  obtained by D. Swartz with the same extraction routines 
  and in the same energy bands. 
Hence, the soft band is defined here
  as the $0.2$--$1.0$ keV channel energy range, 
  instead of the $0.3$--$1.0$ keV band used in the rest of this paper. 
Medium and hard bands are defined as before.

\begin{figure}
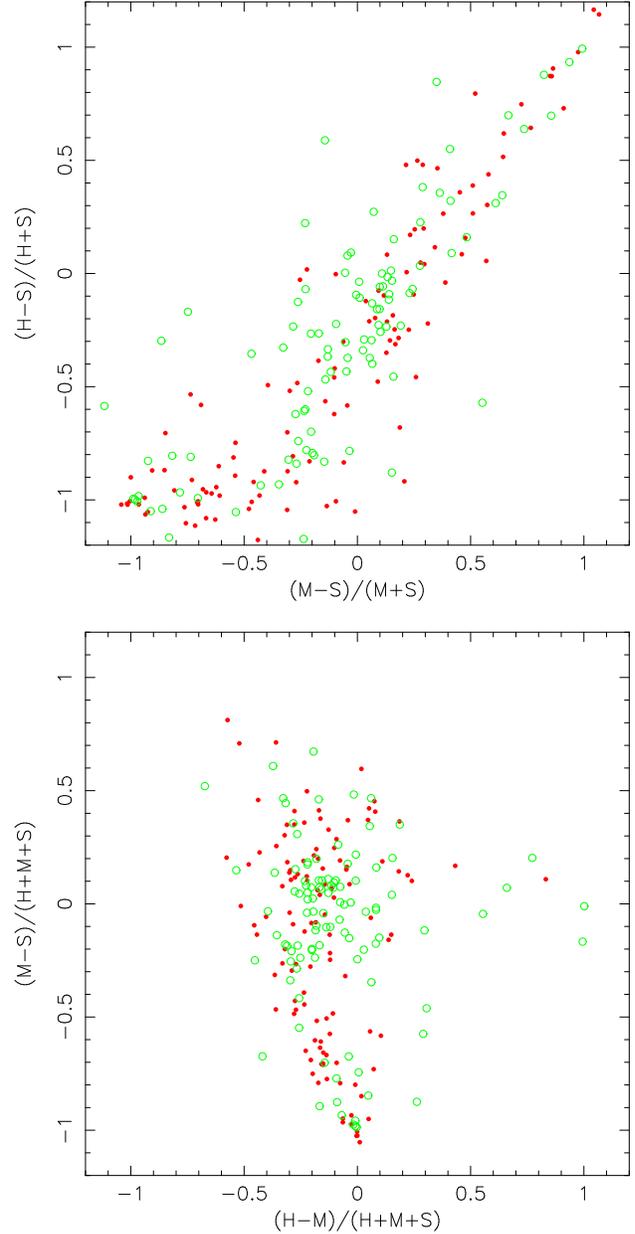

\begin{center} 
\vspace*{0.35cm} 
\epsfig{figure=m83_m81_color_doug.ps, width=8.0cm,angle=270} \\
\vspace{0.35cm} 
\epsfig{figure=m83_m81_color_ap.ps, width=8.0cm,angle=270} \\
\end{center}
\caption{
  Color-color plots for X-ray sources in M\,83 and M\,81, 
     for conventional and alternative color indices (as defined in Sect.~5.2). 
  Here the soft band is defined as $0.2$--$1.0$ keV. 
  Filled red circles represent M\,83 sources, 
    and open green circles represent M\,81 sources. 
Error bars have been omitted. 
}
\label{fig:m83_m81_a}
\end{figure}

\begin{figure}
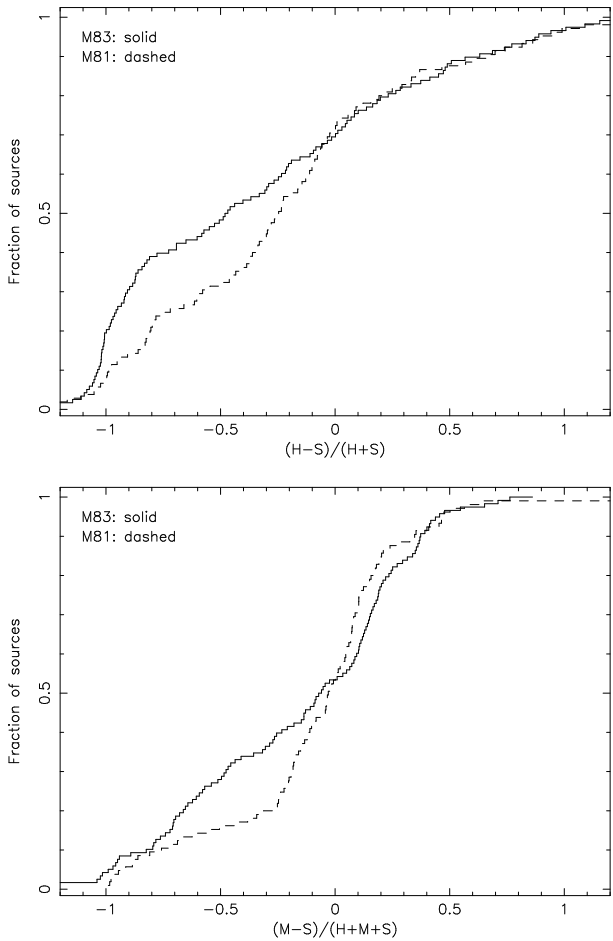

\begin{center} 
\vspace*{0.35cm} 
\epsfig{figure=hardness_distr1.ps, width=6.0cm,angle=270} \\
\vspace{0.35cm} 
\epsfig{figure=hardness_distr3.ps, width=6.0cm,angle=270} \\
\end{center}
\caption{
  The cumulative color distribution of the sources in M\,83 (solid line) 
      and M\,81 (dashed line).  
  S is defined here as the $0.2$--$1.0$ keV band.}
\label{fig:m83_m81_b}
\end{figure}

The color-color plots (Fig.~\ref{fig:m83_m81_a}) and 
  the cumulative hardness distributions 
  (Fig.~\ref{fig:m83_m81_b})
  show similar populations of supersoft sources and XRBs 
  for the two galaxies. 
However, M\,81 has fewer group B sources. 
This difference may be caused 
  either by physical properties 
  (different star-formation activity and history of the two galaxies)  
  or by the higher inclination angle of the M\,81 disk. 
If this group of young soft sources are concentrated in the spiral arms, 
  they may be more highly absorbed and less likely to be detected. 
However, the fact that there is no difference 
  in the relative detection of the other groups of sources 
  suggests that the former explanation is more likely. 
As discussed in Sect.~5.3, 
  a significant fraction of group B sources in M\,83 
  is located in the starburst nuclear region (10 out of 50) 
  and in other young star-forming regions  
  where the interstellar gas density is higher. 
It is natural that fewer of these sources are in the bulge of M\,81, 
  which is characterized by an older stellar population 
  and little interstellar gas.

                          
\section{Summary}  

We have identified 127 discrete sources 
  in a 51-ks {\it Chandra} ACIS-S3 observation 
  of the starburst spiral galaxy M\,83, 
  with a detection limit of $\approx 3 \times 10^{36} {\rm cts~s}^{-1}$ 
  in the $0.3$--$8.0$ keV band.  
Most of the bright sources have spectra typical of XRBs, 
either in a soft state (thermal component at $\sim 1$ keV 
plus powerlaw with $\Gamma \approx 2.5$), or in a hard state 
  (powerlaw with $\Gamma \approx 1.5$). 
Two bright sources show emission lines on a hard powerlaw continuum, 
and are probably XRBs 
  surrounded by a photo-ionized nebula or stellar wind,  
  though the alternative that they may be very young SNRs cannot be ruled out. 
Among the other bright sources, we also modelled the spectra of two SNR candidates, 
with optically-thin thermal plasma emission at temperatures $\sim 0.5$ keV 
and emission lines from Mg and Si. 
The two brightest supersoft sources have blackbody temperatures 
  $kT \approx 60$ eV and luminosities $\sim 10^{38}$ erg s$^{-1}$.
Two candidate X-ray pulsars are detected with periods $\approx 200$~s.

An analysis of the X-ray colors, luminosity and spatial distribution suggests 
  that the discrete source population can be divided into three distinct groups. 
The first group ($\approx 10$\% of the detected sources) comprises 
  ``classical'' supersoft sources with blackbody-like spectra 
at temperatures $< 100$ eV. They do not correlate strongly 
with H$\alpha$ emission, an indicator of recent star formation, 
  suggesting that they belong to an old or intermediate-age stellar population.
This is consistent with their interpretation as nuclear-burning white dwarfs. 
The second group ($\approx 40$\% of the total) 
  consists of soft sources with little or no detected emission above 2 keV. 
The brightest sources in this group 
  have unabsorbed luminosities $\approx 10^{38}$ erg s$^{-1}$, 
  but for most of the others we estimate luminosities 
  $\sim 10^{37}$ erg s$^{-1}$. 
They are mostly found in regions with high H$\alpha$ emission 
(H\,{\footnotesize II} regions in the spiral arms 
and the starburst nucleus). 
Most of them are probably core-collapse SNRs, although other physical systems 
may occasionally show similarly soft X-ray colors (for example, 
accreting neutron stars in high-mass XRBs 
  with large photospheres). 
A comparison with the source colors in the spiral galaxy M\,81 (where 
star formation is currently less active) confirms that 
the fraction of sources detected in this group is  
related to the recent level of star formation.
The sources in the third group are mostly XRBs, reaching higher X-ray 
luminosities than sources in the other two groups. Being a mixture 
of old low-mass and young high-mass XRBs, the whole group appears 
to be of intermediate age when correlated with the H$\alpha$ emission.  
The color-color diagrams allow us to distinguish between 
sources in a soft and hard state, and to disentangle 
the effects of absorption versus intrinsic spectral hardness.

\begin{acknowledgements}

We thank Rosanne Di Stefano, Stefan Immler, Phil Kaaret, 
  Roy Kilgard, Albert Kong, Manfred Pakull, Andrea Prestwich, 
  Doug Swartz, Andreas Zezas for valuable discussions.  
We thank Doug Swartz for providing us 
     his 3-band {\em Chandra} ACIS count rates for sources in M\,81 and M\,83, 
     so that we could carry out a statistical comparison  
     of two populations with an identical data-reduction procedure.
  We also thank: Darragh O'Donoghue for the use of his {\footnotesize EAGLE} 
     Fourier transform code;  
  Rainer Beck and Stuart Ryder for the radio and H$\alpha$ images 
     respectively;
  the Aspen Center for Physics, 
      where part of this work was carried out,
     for the support and hospitality. 
We are grateful to the referee (Wolfgang Pietsch) for useful suggestions that made 
this paper much clearer.
     
\end{acknowledgements}


\appendix 

\section{Discrete sources in Chandra ACIS S3 image of M~83}  

   \begin{table*}
      \caption[]{
    M\,83 sources detected at the 3-$\sigma$ level
    in the $0.3$--$8.0$ keV band, and respective net fluxes in a soft, medium and hard band.  
    Counts in a given band are given in brackets when the source was detected 
    at less than 3-$\sigma$ level in that band. The exposure time was 50.851 ks. 
The {\it ROSAT} cross-identification refers to the HRI sources listed by Immler et al. (1999); 
SW02 is the cross-identification with Soria \& Wu (2002).}
         \label{}
   $$
         \begin{array}[width=textwidth]{rcccrrrrrr}
            \hline
	    \hline
            \noalign{\smallskip}
           {\mathrm{No.}} & {\mathrm{CXOU~Name}} &
          {\mathrm{R.A.(2000)}}     & {\mathrm{Dec.(2000)}}
                 & F_{0.3-8}{\mathrm{~(cts)}}
		 & F_{0.3-1}{\mathrm{~(cts)}}
		 & F_{1-2}{\mathrm{~(cts)}}
		 & F_{2-8}{\mathrm{~(cts)}}
                 & ROSAT & {\mathrm{SW02}}\\
            \noalign{\smallskip}
            \hline
            \noalign{\smallskip}
  1& {\rm ~J}133640.8-295118 & ~~13~36~40.89~& -29~51~18.2~&    23.5\pm08.5     & 18.8\pm05.3 &	(3.0\pm2.0)    & (0.0\pm0.0)& \\  	   
  2& {\rm ~J}133641.0-295110 & ~~13~36~41.01~& -29~51~10.7~&    67.7\pm10.2     & 22.8\pm07.3 &	40.1\pm07.4	& (13.0\pm5.2)& & 1\\   	   
  3& {\rm ~J}133641.4-295045 & ~~13~36~41.41~& -29~50~45.1~&   294.0\pm19.8     & 49.7\pm08.2 &	126.7\pm12.8	& 89.2\pm11.2 & & 2 \\
  4& {\rm ~J}133643.1-295254 & ~~13~36~43.16~& -29~52~54.3~&    10.4\pm03.7     & (4.0\pm2.0)  &	(5.0\pm3.0)	& (0.5\pm0.5)& \\   	   	   
  5& {\rm ~J}133643.5-295107 & ~~13~36~43.53~& -29~51~07.2~&   324.7\pm20.3     & 29.7\pm06.1 & 	120.0\pm12.2	& 121.3\pm12.5 & {\rm H08} & 3\\
  6& {\rm ~J}133648.0-295324 & ~~13~36~48.05~& -29~53~24.2~&    60.8\pm08.9     & 13.2\pm04.2 &	23.2\pm05.1	& 13.0\pm04.0 & & 4\\
  7& {\rm ~J}133648.2-295244 & ~~13~36~48.27~& -29~52~44.8~&    19.3\pm05.0     & 10.4\pm03.6 &	(6.0\pm3.0)	& (0.5\pm0.5)& \\   	  
  8& {\rm ~J}133649.1-295258 & ~~13~36~49.11~& -29~52~58.5~&   317.5\pm18.8    & 189.8\pm14.5 &	109.7\pm11.0	& 15.2\pm04.2 & {\rm H12a} & 5\\
  9& {\rm ~J}133649.2-295303 & ~~13~36~49.22~& -29~53~03.5~&   112.9\pm11.4     & 35.9\pm06.6 &	53.7\pm07.7	& 12.6\pm03.9 & {\rm H12b} & 6\\
 10& {\rm ~J}133649.8-295217 & ~~13~36~49.82~& -29~52~17.7~&    38.2\pm07.2     & 28.2\pm06.1 &	(8.1\pm3.2)	& (0.5\pm0.5)& & 7\\	  
 11& {\rm ~J}133650.8-295240 & ~~13~36~50.87~& -29~52~40.5~&    27.1\pm06.4     & 20.6\pm05.4 & 	(4.0\pm2.0)	& (0.5\pm0.5)& \\   	     	   
 12& {\rm ~J}133650.9-295226 & ~~13~36~50.97~& -29~52~26.3~&    21.8\pm05.8     & 14.7\pm04.7 &	(4.0\pm2.0)	& (0.5\pm0.5)& \\   	   	   
 13& {\rm ~J}133651.1-295043 & ~~13~36~51.15~& -29~50~43.2~&    19.2\pm05.4     & 10.6\pm03.9 &	(6.0\pm3.0)	& (0.5\pm0.5)& \\   	   	   
 14& {\rm ~J}133651.6-295335 & ~~13~36~51.65~& -29~53~35.5~&   124.0\pm12.0     & 42.6\pm07.1 &	52.4\pm07.5	& 24.7\pm05.2 & {\rm H14} & 8\\
 15& {\rm ~J}133652.8-295231 & ~~13~36~52.80~& -29~52~31.4~&    32.1\pm06.7     & 20.6\pm05.4 & 	10.8\pm03.5 	& (0.5\pm0.5)& \\	   	   
 16& {\rm ~J}133653.1-295430 & ~~13~36~53.19~& -29~54~30.4~&    27.5\pm06.0     & (5.0\pm2.0) &	15.2\pm04.0 	& (5.5\pm2.5)& \\	   	   
 17& {\rm ~J}133653.1-295325 & ~~13~36~53.19~& -29~53~25.3~&    43.1\pm07.7     & 34.5\pm06.6 & 	(5.5\pm3.2) 	& (3.2\pm2.5)& & 9\\  	   
 18& {\rm ~J}133653.2-295242 & ~~13~36~53.27~& -29~52~42.9~&    46.3\pm08.2     & 19.1\pm05.3 & 	16.0\pm04.2	& (3.5\pm2.9)& & 10\\   	   
 19& {\rm ~J}133653.6-295600 & ~~13~36~53.65~& -29~56~00.9~&    30.0\pm06.0     & 17.5\pm04.5 &	(6.7\pm2.8)	& (2.7\pm1.6)& \\   	   
 20& {\rm ~J}133653.9-294848 & ~~13~36~53.92~& -29~48~48.8~&    32.6\pm06.9     & 28.1\pm06.2 &	 (4.0\pm2.0)  	& (0.5\pm0.5)& & 11\\   	   
 21& {\rm ~J}133653.9-295114 & ~~13~36~53.93~& -29~51~14.7~&    89.5\pm10.2     & 43.7\pm07.1 &	30.6\pm05.7	& 13.1\pm03.9 & & 12\\
 22& {\rm ~J}133654.1-295209 & ~~13~36~54.17~& -29~52~09.6~&    12.3\pm04.1     & 10.5\pm03.5  &	 (2.0\pm1.0)  	& (0.5\pm0.5)& \\   	   
 23& {\rm ~J}133654.1-294933 & ~~13~36~54.18~& -29~49~33.1~&    28.8\pm07.5     & (12.6\pm5.6)&	(12.5\pm5.2)   	& (3.8\pm2.0)& \\   	   
 24& {\rm ~J}133655.0-295304 & ~~13~36~55.06~& -29~53~04.9~&    25.5\pm05.9     & 15.9\pm04.6 &	11.2\pm03.7	& (0.0\pm0.0)& \\   	   
 25& {\rm ~J}133655.0-295239 & ~~13~36~55.09~& -29~52~39.6~&    29.6\pm06.2     & 24.2\pm05.4 &	(4.0\pm2.0)   	& (0.5\pm0.5)& & 13\\   	   
 26& {\rm ~J}133655.2-295403 & ~~13~36~55.23~& -29~54~03.6~&    98.4\pm10.5     & 41.7\pm07.1 &	36.1\pm06.1	& 24.4\pm05.1 & & 14\\
 27& {\rm ~J}133655.5-295510 & ~~13~36~55.50~& -29~55~10.1~&   352.3\pm19.2     & 94.7\pm10.1 &	151.3\pm12.5	& 108.6\pm10.7 & {\rm H15} & 15\\
 28& {\rm ~J}133655.6-295303 & ~~13~36~55.60~& -29~53~03.6~&    48.1\pm07.9     & 23.5\pm06.1 &	8.7\pm03.2	& 15.6\pm04.1 & & 16\\
 29& {\rm ~J}133656.2-295255 & ~~13~36~56.22~& -29~52~55.5~&    22.9\pm05.5     & (5.5\pm2.3) &	12.3\pm03.7	& (6.8\pm2.6)& & 17\\   	   
 30& {\rm ~J}133656.5-294817 & ~~13~36~56.51~& -29~48~17.7~&    21.5\pm04.1     & (2.0\pm1.0) &	11.3\pm03.5	& (7.0\pm3.0)& \\   	   
 31& {\rm ~J}133656.6-294912 & ~~13~36~56.65~& -29~49~12.6~&   311.6\pm18.8     & 53.3\pm07.7 &	145.0\pm12.7	& 95.1\pm10.5 & & 18\\
 32& {\rm ~J}133656.6-295321 & ~~13~36~56.69~& -29~53~21.8~&    14.3\pm04.5     & (11.0\pm4.0)&	(1.0\pm1.0)   	& (2.0\pm1.0)&\\   	   
 33& {\rm ~J}133657.2-295339 & ~~13~36~57.28~& -29~53~39.7~&   490.6\pm22.7    & 162.8\pm13.3 &	201.6\pm14.4	& 126.8\pm11.4 &  & 19\\
 34& {\rm ~J}133657.5-294729 & ~~13~36~57.52~& -29~47~29.2~&    96.4\pm12.5     & 35.1\pm08.1  &	42.5\pm07.6	& 32.6\pm07.6 & & 20\\
 35& {\rm ~J}133657.8-295303 & ~~13~36~57.89~& -29~53~03.0~&    81.0\pm09.4     & 61.7\pm08.1 &	18.6\pm04.5	& (0.5\pm0.5)& & 21\\   	   
 36& {\rm ~J}133657.9-294923 & ~~13~36~57.94~& -29~49~23.6~&    35.2\pm06.7     & (1.0\pm1.0)&	12.3\pm03.7	& 16.9\pm04.4 & & 22\\
 37& {\rm ~J}133658.2-295124 & ~~13~36~58.20~& -29~51~24.6~&    20.7\pm05.3     & (15.5\pm6.2) &	10.6\pm03.6	& (3.0\pm2.0)& & 23\\   	   
 38& {\rm ~J}133658.2-295216 & ~~13~36~58.23~& -29~52~16.0~&   164.3\pm13.4     & 91.9\pm10.0 &	63.0\pm08.1	& 12.3\pm03.7 & & 24\\
^{\mathrm{a}} 39& {\rm ~J}133658.3-295105 & ~~13~36~58.39~& -29~51~05.1~&   124.9\pm11.8    & (2.0\pm1.0) &	54.1\pm07.6	& 70.4\pm08.6 & & 26\\
 40& {\rm ~J}133658.4-294833 & ~~13~36~58.41~& -29~48~33.4~&    76.1\pm10.2    & 26.1\pm07.7  &	21.2\pm05.1	& 35.2\pm06.9 & & 25\\

   \noalign{\smallskip}
            \hline
         \end{array}
     $$
\begin{list}{}{}
\item[$^{\mathrm{a}}$] Background radio galaxy 
\end{list}
\end{table*}

\begin{table*}
         \label{}
   $$
         \begin{array}[width=textwidth]{rcccrrrrrr}
            \hline
	    \hline
            \noalign{\smallskip}
           {\mathrm{No.}} & {\mathrm{CXOU~Name}} &
          {\mathrm{R.A.(2000)}}     & {\mathrm{Dec.(2000)}}
                 & F_{0.3-8}{\mathrm{~(cts)}}
		 & F_{0.3-1}{\mathrm{~(cts)}}
		 & F_{1-2}{\mathrm{~(cts)}}
		 & F_{2-8}{\mathrm{~(cts)}}
                 & ROSAT & {\mathrm{SW02}}\\
            \noalign{\smallskip}
            \hline
            \noalign{\smallskip}

 41& {\rm ~J}133658.7-295100 & ~~13~36~58.76~& -29~51~00.9~&    16.0\pm04.8     & 17.9\pm05.1 &	(0.5\pm0.5)   	& (0.5\pm0.5)& \\   	   
 42& {\rm ~J}133659.1-295336 & ~~13~36~59.13~& -29~53~36.5~&    34.1\pm06.5     & 32.8\pm06.2 &	(0.5\pm0.5)   	& (0.5\pm0.5)& & 27\\   	   
 43& {\rm ~J}133659.3-295508 & ~~13~36~59.35~& -29~55~08.9~&    11.4\pm03.9      & 8.0\pm03.2 &	(1.0\pm1.0)   	& (0.5\pm0.5)& \\   	   
 44& {\rm ~J}133659.4-294959 & ~~13~36~59.46~& -29~49~59.3~&  1214.1\pm36.0    & 399.2\pm20.7 &	474.2\pm22.4	& 347.9\pm19.3 & {\rm H17} & 28\\
45& {\rm ~J}133659.5-295203 & ~~13~36~59.52~& -29~52~03.9~&    36.0\pm12.5     & 29.7\pm10.0 &	(9.0\pm3.5)   	&  (0.0\pm0.0)& (^{\mathrm{b}}) & 29\\    
 46& {\rm ~J}133659.5-295413 & ~~13~36~59.59~& -29~54~13.5~&    23.6\pm05.6     & (1.0\pm1.0)&	11.8\pm03.6	& 9.1\pm03.2 & \\
47& {\rm ~J}133659.8-295205 & ~~13~36~59.81~& -29~52~05.5~&   270.4\pm21.0     & 85.7\pm13.9 &	137.6\pm12.7	& 62.2\pm08.2  & (^{\mathrm{b}}) &30\\
 48& {\rm ~J}133659.8-295525 & ~~13~36~59.85~& -29~55~25.9~&    31.1\pm06.2     & 24.2\pm05.1 &	(1.0\pm1.0)   	& (2.0\pm1.0)& \\   	   
49& {\rm ~J}133700.0-295150 & ~~13~37~00.03~& -29~51~50.2~&   192.1\pm22.7     & 77.9\pm15.5 &	79.8\pm10.9	& 48.5\pm07.4  & (^{\mathrm{b}}) & 31\\
 50& {\rm ~J}133700.0-295138 & ~~13~37~00.06~& -29~51~38.1~&    17.7\pm05.8     & (6.0\pm3.0)&	12.3\pm04.1	& (0.5\pm0.5)& \\   	   
 51& {\rm ~J}133700.0-295417 & ~~13~37~00.08~& -29~54~17.5~&    11.7\pm04.0      & 9.7\pm03.6 &	 (1.0\pm1.0)  	& (0.5\pm0.5)& \\   	   
 52& {\rm ~J}133700.1-295330 & ~~13~37~00.11~& -29~53~30.0~&    36.8\pm06.4    & (2.0\pm1.0)	 	&20.7\pm04.6	& 13.1\pm03.7 & & 32\\
53& {\rm ~J}133700.2-295206 & ~~13~37~00.23~& -29~52~06.3~&    41.8\pm11.6      & 28.1\pm10.4 &	19.5\pm6.3     	& (1.7\pm1.7) & (^{\mathrm{b}}) & 33\\   
54& {\rm ~J}133700.3-295205 & ~~13~37~00.36~& -29~52~05.6~&    37.1\pm12.1     & 24.9\pm09.2 &	(6.2\pm4.9)    	& (0.5\pm0.5) & (^{\mathrm{b}}) & 34\\    
 55& {\rm ~J}133700.4-295054 & ~~13~37~00.42~& -29~50~54.3~&    94.1\pm10.4     & 91.2\pm10.1 &	(0.5\pm0.5)   	& (0.0\pm0.0)& & 36\\   	   
56& {\rm ~J}133700.4-295159 & ~~13~37~00.45~& -29~51~59.5~&   325.4\pm37.8     & 290.4\pm33.5 &	45.0\pm21.9	& (3.2\pm3.2)& (^{\mathrm{b}}) & 35\\    
57& {\rm ~J}133700.5-295156 & ~~13~37~00.51~& -29~51~56.1~&   294.0\pm39.7     & 220.9\pm31.5 &	77.0\pm24.2	& (4.8\pm4.8) & (^{\mathrm{b}}) & 37\\    
58& {\rm ~J}133700.6-295146 & ~~13~37~00.60~& -29~51~46.7~&    63.9\pm13.0     & (8.1\pm4.9)&	26.8\pm06.6	& 17.4\pm04.6  & (^{\mathrm{b}}) & 38\\
59& {\rm ~J}133700.6-295159 & ~~13~37~00.65~& -29~51~59.5~&   181.1\pm39.0     & 112.6\pm35.0 &	45.0\pm14.7	&  (2.0\pm2.0)& (^{\mathrm{b}}) & 40\\    
 60& {\rm ~J}133700.6-295319 & ~~13~37~00.65~& -29~53~19.9~&   395.0\pm20.2    & 125.6\pm11.5 &	160.4\pm12.8	& 110.2\pm10.6 & & 39\\
61& {\rm ~J}133700.6-295204 & ~~13~37~00.66~& -29~52~04.0~&   115.1\pm32.4     & 85.7\pm21.0 &	(32.3\pm13.7)   & (0.5\pm0.5) & (^{\mathrm{b}}) & 41\\    
62& {\rm ~J}133700.7-295205 & ~~13~37~00.72~& -29~52~05.9~&   159.3\pm30.5     & (18.2\pm8.8)&	99.2\pm13.0	& 66.0\pm08.7  & (^{\mathrm{b}}) & 42\\
63& {\rm ~J}133700.9-295155 & ~~13~37~00.90~& -29~51~55.7~&   739.4\pm36.6    & 204.6\pm24.8 &	274.4\pm20.9	& 252.9\pm16.6  & (^{\mathrm{b,c}}) & 43\\
64& {\rm ~J}133700.9-295202 & ~~13~37~00.97~& -29~52~02.8~&  1327.1\pm45.7    & 570.9\pm33.4 &	589.5\pm27.2	& 164.8\pm13.4  & (^{\mathrm{b}}) & 44\\
 65& {\rm ~J}133701.0-295245 & ~~13~37~01.07~& -29~52~45.8~&   113.4\pm11.2     & 51.3\pm07.6 &	39.5\pm06.5	& 25.8\pm05.2 & {\rm H21} & 45\\
 66& {\rm ~J}133701.0-295056 & ~~13~37~01.08~& -29~50~56.2~&    22.4\pm05.9     & 21.8\pm05.5 &	(1.0\pm1.0)   	& (0.5\pm0.5)& \\   	   
67& {\rm ~J}133701.1-295151 & ~~13~37~01.16~& -29~51~51.8~&    81.4\pm15.4     & (14.0\pm9.1) &   47.5\pm09.0	& 14.0\pm04.4  & (^{\mathrm{b}}) & 46\\
 68& {\rm ~J}133701.1-295449 & ~~13~37~01.18~& -29~54~49.5~&   123.5\pm11.4    & 123.4\pm11.4 &	(0.5\pm0.5)   	& (0.0\pm0.0)& & 47\\  	   
69& {\rm ~J}133701.2-295201 & ~~13~37~01.27~& -29~52~01.7~&    73.0\pm11.7     & (9.0\pm7.5)	&(13.0\pm5.4)	& 51.4\pm07.9  & (^{\mathrm{b}}) & 48\\
70& {\rm ~J}133701.2-295159 & ~~13~37~01.28~& -29~51~59.9~&    103.0\pm25.2     & 69.2\pm10.0 &	40.1\pm09.4	& (5.0\pm2.0) & (^{\mathrm{b}}) & 49\\   	   
 71& {\rm ~J}133701.3-295136 & ~~13~37~01.31~& -29~51~36.9~&    54.8\pm08.0    & (1.0\pm1.0) & (3.0\pm2.0)	& 54.8\pm07.6 & & 50\\
 72& {\rm ~J}133701.4-295326 & ~~13~37~01.49~& -29~53~26.7~&   488.9\pm22.5     & 75.1\pm08.8 &	307.9\pm17.7	& 103.3\pm10.4 & {\rm H20} & 51\\
 73& {\rm ~J}133701.6-295128 & ~~13~37~01.64~& -29~51~28.4~&   535.7\pm23.9    & 224.7\pm15.6 &	213.3\pm14.9	& 97.4\pm10.1 & {\rm H23} & 52\\
 74& {\rm ~J}133701.6-295201 & ~~13~37~01.65~& -29~52~01.9~&    22.9\pm06.8     & 21.6\pm06.4 &	(2.0\pm1.0)   	&(0.5\pm0.5)  & (^{\mathrm{b}})\\   
 75& {\rm ~J}133701.6-295410 & ~~13~37~01.68~& -29~54~10.1~&    12.8\pm04.1     & (9.0\pm4.0) &	(2.0\pm1.0)   	& (0.0\pm0.0)& \\   	   
 76& {\rm ~J}133701.7-294743 & ~~13~37~01.72~& -29~47~43.0~&   173.2\pm15.1     & 14.3\pm04.4 &	71.7\pm09.5	& 48.5\pm07.9 & & 53\\
 77& {\rm ~J}133701.7-295113 & ~~13~37~01.72~& -29~51~13.5~&    27.1\pm06.1     & 20.6\pm05.1 &	 (3.0\pm2.0)  	& (0.5\pm0.5)& & 54\\   	   
 78& {\rm ~J}133702.0-295518 & ~~13~37~02.03~& -29~55~18.0~&   333.8\pm18.6    & 112.2\pm10.9 &	96.5\pm09.9	& 125.7\pm11.4 & {\rm H22} & 55\\
 79& {\rm ~J}133702.1-295505 & ~~13~37~02.16~& -29~55~05.9~&    54.4\pm07.9     & 17.0\pm04.5 &	19.5\pm04.6	& 10.4\pm03.5 & & 56\\
 80& {\rm ~J}133702.2-294952 & ~~13~37~02.23~& -29~49~52.9~&    27.6\pm05.8     & 18.2\pm04.7 &	6.3\pm02.6	& (3.0\pm2.0)& & 57\\   	   
 81& {\rm ~J}133702.4-295126 & ~~13~37~02.43~& -29~51~26.3~&    78.7\pm09.5     & 60.6\pm08.3 &	17.4\pm04.5	& (0.5\pm0.5)& & 58\\   	   
 82& {\rm ~J}133702.4-295319 & ~~13~37~02.47~& -29~53~19.4~&    97.4\pm10.1     & 13.9\pm03.9 &	51.9\pm07.3	& 31.5\pm05.7 & & 59\\
 83& {\rm ~J}133703.0-294945 & ~~13~37~03.06~& -29~49~45.4~&    18.7\pm05.2     & (10.6\pm6.2)&	7.0\pm02.8	& (0.5\pm0.5)& \\   	   
 84& {\rm ~J}133703.3-295226 & ~~13~37~03.31~& -29~52~26.9~&    64.9\pm08.6     & 14.7\pm04.4 &	27.8\pm05.4	& 23.7\pm05.0 & & 60\\
 85& {\rm ~J}133703.8-294930 & ~~13~37~03.88~& -29~49~30.6~&   346.0\pm19.5    & 114.0\pm11.3 &	141.7\pm12.4	& 89.4\pm09.8 & & 61\\
 86& {\rm ~J}133704.2-295403 & ~~13~37~04.29~& -29~54~03.8~&  1389.2\pm37.8    & 406.8\pm20.4 &	702.4\pm26.8	& 283.4\pm17.0 & {\rm H26}& 62\\
 87& {\rm ~J}133704.3-295130 & ~~13~37~04.38~& -29~51~30.7~&    41.0\pm07.0     & 16.2\pm04.5 &	19.4\pm04.6	& (5.0\pm2.0)& & 63\\   	   
 88& {\rm ~J}133704.3-295121 & ~~13~37~04.38~& -29~51~21.7~&  1527.8\pm39.7    & 469.6\pm22.0 &	762.6\pm27.9	& 302.0\pm17.6 & {\rm H27a} & 64\\
 89& {\rm ~J}133704.4-294938 & ~~13~37~04.46~& -29~49~38.8~&    12.1\pm04.1      & 7.8\pm03.3 &	(2.0\pm1.0)   	& (1.0\pm1.0)& \\   	   
 90& {\rm ~J}133704.5-295108 & ~~13~37~04.59~& -29~51~08.3~&    13.2\pm04.4     & 10.6\pm03.9 &	(2.0\pm1.0)   	& (0.5\pm0.5)& \\

   \noalign{\smallskip}
            \hline
         \end{array}
     $$
\begin{list}{}{}
\item[$^{\mathrm{b}}$] Contributing to the nuclear source H19, unresolved by {\it ROSAT}
\item[$^{\mathrm{c}}$] X-ray source coincident with the IR/optical/UV nucleus of the galaxy
\end{list}
   \end{table*}

  \begin{table*}
         \label{}
   $$
         \begin{array}[width=textwidth]{rcccrrrrrr}
            \hline
	    \hline
            \noalign{\smallskip}
           {\mathrm{No.}} &  {\mathrm{CXOU~Name}} &
          {\mathrm{R.A.(2000)}}     & {\mathrm{Dec.(2000)}}
                 & F_{0.3-8}{\mathrm{~(cts)}}
		 & F_{0.3-1}{\mathrm{~(cts)}}
		 & F_{1-2}{\mathrm{~(cts)}}
		 & F_{2-8}{\mathrm{~(cts)}}
                 & ROSAT & {\mathrm{SW02}}\\
            \noalign{\smallskip}
            \hline
            \noalign{\smallskip}

 91& {\rm ~J}133704.6-295120 & ~~13~37~04.66~& -29~51~20.6~&    69.8\pm09.0     & 20.6\pm05.0 &	31.0\pm06.0	& 19.6\pm04.7 & {\rm H27b} & 65\\
 92& {\rm ~J}133704.7-294851 & ~~13~37~04.77~& -29~48~51.9~&    37.9\pm07.0     & (10.6\pm4.5)&	16.2\pm04.2	& (6.7\pm3.9)& & 66\\   	   
 93& {\rm ~J}133705.4-295234 & ~~13~37~05.48~& -29~52~34.2~&    89.9\pm09.8    & (1.0\pm1.0)  &	39.8\pm06.4	& 51.3\pm07.3 & & 67\\
 94& {\rm ~J}133706.0-295159 & ~~13~37~06.00~& -29~51~59.5~&    17.8\pm04.6     & (1.0\pm1.0) &	9.3\pm03.2	& 7.3\pm02.8 & \\
 95& {\rm ~J}133706.0-295514 & ~~13~37~06.04~& -29~55~14.7~&    72.6\pm08.9     & 53.2\pm07.5 &	15.3\pm04.0	& (1.0\pm1.0)& & 68\\   	   
 96& {\rm ~J}133706.1-295232 & ~~13~37~06.18~& -29~52~32.4~&   119.8\pm11.5    & 110.9\pm10.9 &	(2.3\pm2.0)   	& (0.5\pm0.5)& & 70\\   	   
 97& {\rm ~J}133706.1-295444 & ~~13~37~06.19~& -29~54~44.3~&    42.3\pm06.9     & 36.1\pm06.3 &	6.7\pm02.6	& (0.5\pm0.5)& & 69\\   	   
 98& {\rm ~J}133706.5-295416 & ~~13~37~06.54~& -29~54~16.0~&    15.9\pm04.5     & 15.3\pm04.2 &	(0.5\pm0.5)   	& (0.0\pm0.0)& \\   	   
 99& {\rm ~J}133706.6-295107 & ~~13~37~06.63~& -29~51~07.6~&    13.7\pm04.4     & (3.0\pm2.0) &	4.3\pm02.2	& 6.0\pm02.0& \\   	   
100& {\rm ~J}133706.6-295332 & ~~13~37~06.67~& -29~53~32.8~&    25.4\pm05.7      & 8.2\pm03.2 &	9.8\pm03.3	& (5.0\pm2.0)& \\   	   
101& {\rm ~J}133706.8-295057 & ~~13~37~06.81~& -29~50~57.8~&    17.7\pm05.6     & (8.3\pm5.8) &	(2.0\pm1.0)   	& (6.0\pm4.0)& \\   	   
102& {\rm ~J}133707.0-295320 & ~~13~37~07.06~& -29~53~20.8~&    11.4\pm04.1     & (10.0\pm4.0)&	(1.0\pm1.0)   	& (0.5\pm0.5)& \\   	   
103& {\rm ~J}133707.1-295202 & ~~13~37~07.10~& -29~52~02.3~&    17.9\pm04.7     & 11.5\pm03.7 &	(4.0\pm2.0)   	& (1.0\pm1.0)& \\   	   
104& {\rm ~J}133707.1-295101 & ~~13~37~07.12~& -29~51~01.8~&   762.4\pm28.2     & 22.5\pm05.4 &	325.7\pm18.3	& 418.8\pm20.8 & & 71\\
105& {\rm ~J}133707.4-295133 & ~~13~37~07.47~& -29~51~33.9~&    16.8\pm04.8     & 15.7\pm04.5 &	(0.5\pm0.5)   	& (0.5\pm0.5)& \\   	   
106& {\rm ~J}133707.5-294859 & ~~13~37~07.53~& -29~48~59.5~&    22.9\pm05.5     & 19.1\pm04.9 &	(2.0\pm1.0)   	& (1.0\pm1.0)& \\   	   
107& {\rm ~J}133708.2-294916 & ~~13~37~08.21~& -29~49~16.6~&    13.6\pm04.5     & (12.0\pm5.0)&	(1.0\pm1.0)   	& (1.0\pm1.0)& \\   	   
108& {\rm ~J}133708.3-295126 & ~~13~37~08.37~& -29~51~26.2~&    21.7\pm05.7     & 13.3\pm04.5 &	(4.0\pm2.0)   	& (1.0\pm1.0)& \\   	   
109& {\rm ~J}133708.5-295135 & ~~13~37~08.57~& -29~51~35.2~&    11.4\pm04.1      & 9.2\pm03.5 &	 (1.0\pm1.0)   	& (0.0\pm0.0)& \\   	   
110& {\rm ~J}133711.9-295215 & ~~13~37~11.90~& -29~52~15.8~&    56.1\pm08.1     & 45.5\pm07.1 &	8.3\pm03.0	& (0.5\pm0.5)& & 72\\   	   
111& {\rm ~J}133712.1-295056 & ~~13~37~12.10~& -29~50~56.4~&    18.7\pm05.3     & 15.8\pm04.7 &	(1.0\pm1.0)   	& (2.0\pm1.0)& \\   	   
112& {\rm ~J}133712.4-295140 & ~~13~37~12.46~& -29~51~40.0~&    28.9\pm05.7     & 13.7\pm03.9 &	14.5\pm03.9	& (1.0\pm1.0)& & 73\\   	   
113& {\rm ~J}133712.5-295155 & ~~13~37~12.52~& -29~51~55.0~&   177.9\pm13.8    & (1.0\pm1.0)  &	69.2\pm08.5	& 104.9\pm10.3 & & 74\\
114& {\rm ~J}133712.7-295201 & ~~13~37~12.74~& -29~52~01.3~&     7.4\pm03.2      & 6.4\pm02.8 &	(0.0\pm0.0)   	& (0.5\pm0.5)& \\   	   
115& {\rm ~J}133712.8-295012 & ~~13~37~12.84~& -29~50~12.4~&    40.0\pm06.7     & 37.3\pm06.4 &	(0.5\pm0.5)   	& (0.5\pm0.5)& & 75\\   	   
116& {\rm ~J}133713.1-295238 & ~~13~37~13.14~& -29~52~38.8~&    14.0\pm04.4     & (0.5\pm0.5) &	(5.0\pm2.0)   	& (7.0\pm2.0)& \\   	   
117& {\rm ~J}133714.4-295149 & ~~13~37~14.47~& -29~51~49.1~&   105.9\pm10.5     & 17.7\pm04.4 &	56.3\pm07.6	& 32.6\pm05.8 & & 76\\
118& {\rm ~J}133714.6-294944 & ~~13~37~14.69~& -29~49~44.3~&    55.2\pm07.7     & 14.1\pm03.9 &	24.7\pm05.2	& 15.6\pm04.1 & & 77\\
119& {\rm ~J}133714.7-295428 & ~~13~37~14.75~& -29~54~28.3~&    14.5\pm03.9     & (2.0\pm1.0) &	6.0\pm02.4	& 4.9\pm02.2 & \\
120& {\rm ~J}133716.1-295202 & ~~13~37~16.18~& -29~52~02.4~&    14.9\pm04.1     & (1.0\pm1.0) &	4.9\pm02.2	& 9.0\pm03.2 & \\
121& {\rm ~J}133716.3-294939 & ~~13~37~16.38~& -29~49~39.5~&   480.6\pm23.0    & 112.1\pm11.0 &	199.5\pm14.9	& 174.0\pm13.9 & {\rm H29} & 78\\
122& {\rm ~J}133717.2-295153 & ~~13~37~17.22~& -29~51~53.6~&    91.5\pm10.1     & 61.6\pm08.2 &	28.6\pm05.6	& (2.0\pm1.0)& & 79\\   	   
123& {\rm ~J}133717.4-295154 & ~~13~37~17.48~& -29~51~54.2~&    21.1\pm07.2     & (12.0\pm5.0)&	(6.0\pm3.0)   	& (1.0\pm1.0)& \\   	   
124& {\rm ~J}133717.9-295211 & ~~13~37~17.97~& -29~52~11.9~&    40.3\pm06.6     & 21.4\pm04.8 &	10.9\pm03.3	& (5.0\pm2.0)& & 80\\   	   
125& {\rm ~J}133718.8-295014 & ~~13~37~18.86~& -29~50~14.4~&    10.7\pm03.7     & (1.0\pm1.0) &	5.5\pm02.4	& (1.0\pm1.0)& \\   	   
126& {\rm ~J}133719.6-295131 & ~~13~37~19.65~& -29~51~31.6~&    83.5\pm09.5    & (4\pm2)	&	20.0\pm04.6	& 58.5\pm07.9 & & 81\\
127& {\rm ~J}133722.1-295207 & ~~13~37~22.13~& -29~52~07.8~&    11.6\pm03.7    & (1.0\pm1.0)  &	5.8\pm02.4	& (2.0\pm1.0)& \\

   \noalign{\smallskip}
            \hline
         \end{array}
     $$
   \end{table*}



\clearpage 

\section{Fit parameters of selected bright sources} 

\begin{table}
\caption{
   {\small XSPEC} best-fit parameters 
   for the galactic nucleus (No.~63 in Table A.1).}
\label{mathmode} 
\centering 
\begin{tabular}{@{}lrr} 
\hline
\hline \\ 
\multicolumn{3}{c}{model: wabs$_{\rm Gal}$ $\times$ wabs $\times$ powerlaw  }\\[5pt]  
\hline \\ 
$n_{\rm H}~(\times 10^{21}$~cm$^{-2})$  & $1.25^{+0.80}_{-0.93}$   \\[5pt] 
$\Gamma$       & $1.45^{+0.16}_{-0.24}$ \\[5pt]
$K_{\rm pl}~(\times 10^{-5})$    & $1.9^{+0.3}_{-0.5}$ \\ [5pt]   
\hline \\ 
$\chi_\nu^2$~(dof) &  1.06~(49)  \\ [5pt] 
$L_{\rm{0.3-8}}~(\times 10^{38}$~erg~s$^{-1})$ & $2.3^{+0.2}_{-0.1}$  \\[5pt]
\hline \\
\end{tabular}     
\end{table}



\begin{table}
\caption{
   {\small XSPEC} best-fit parameters 
   for the two brightest supersoft sources. (An asterisk indicates 
that we could not determine an error for that fit parameter).}
\centering 
\begin{tabular}{@{}lrr}
\hline
\hline \\ 
parameter &  No.~68 & No.~96 \\  [5pt]
\hline
\hline \\ 
\multicolumn{3}{c}{model: wabs$_{\rm Gal}$ $\times$ wabs $\times$ blackbody  }\\[5pt]  
\hline \\
$n_{\rm H}~(\times 10^{21}$~cm$^{-2})$  & $1.4^{+1.4}_{-0.5}$  & $4.7^{+2.3}_{-1.0}$ \\[5pt] 
$T_{\rm bb}$~(keV)       & $0.065^{+0.014}_{-0.013}$ & $0.058^{+0.032}_{-0.023}$\\[5pt]
$K_{\rm bb}~(\times 10^{-6})$    & $2.3^{+0.3}_{-0.4}$ & $48^{+35}_{-25}$\\ [5pt]   
\hline \\ 
$\chi_\nu^2$~(dof) &  1.01~(7) &  0.51~(5) \\ [5pt] 
$L_{\rm{0.3-8}}~(\times 10^{38}$~erg~s$^{-1})$ & $0.9^{+3.4}_{-0.4}$ & [$> 1$]  \\[5pt]
\hline
\hline \\ 
\multicolumn{3}{c}{model: wabs$_{\rm Gal}$ $\times$ wabs $\times$ raymond-smith  }\\[5pt]  
\hline \\
$n_{\rm H}~(\times 10^{21}$~cm$^{-2})$  & $1.9^{+1.5}_{-1.2}$  & $4.5^{+150}_{-4.5}$ \\[5pt] 
$T_{\rm rs}$~(keV)       & $0.087^{+0.033}_{-0.027}$ & $0.082^{+0.007}_{-0.013}$\\[5pt]
Z (metal ab.) & $0.01^{+0.04}_{-0.01}$ & $0.02^{+0.14}_{-0.02}$  \\[5pt]
$K_{\rm rs}~(\times 10^{-3})$    & $7.7^{+8.0}_{-4.5}$ & $59^{+*}_{-*}$\\ [5pt]   
\hline \\ 
$\chi_\nu^2$~(dof) &  1.08~(6) &  0.41~(4) \\ [5pt] 
$L_{\rm{0.3-8}}~(\times 10^{38}$~erg~s$^{-1})$ & $2.0^{+10.2}_{-1.5}$ & [$> 1$]  \\[5pt]
\hline \\  
\end{tabular}  
\end{table}
 


\begin{table*}
\caption{
   {\small XSPEC} best-fit parameters 
   for emission-line sources. (Parameters listed as 
``undetermined'' are those for which we could not 
obtain meaningful values with {\small XSPEC}.)}
\label{mathmode} 
\centering 
\begin{tabular}{@{}lrrrr} 
\hline
\hline \\
  parameter &  No.~3 & No.~8 & No.~27 & No.~56 \\ [5pt] 	
\hline
\hline \\ 
\multicolumn{5}{c}{model: wabs$_{\rm Gal}$ $\times$ wabs $\times$ powerlaw  }\\[5pt] 
\hline \\
$n_{\rm H}~(\times 10^{21}$~cm$^{-2})$  & $1.3^{+0.8}_{-0.9}$ & $2.2^{+0.9}_{-0.6}$ 
		 & $0.7^{+0.6}_{-0.7}$  &  $>104$\\[3pt] 
$\Gamma$       & $1.51^{+0.30}_{-0.16}$ & $3.64^{+0.54}_{-0.35}$ 
		& $1.35^{+0.14}_{-0.15}$ &  $>10$     \\[3pt]   
$K_{\rm pl}~(\times 10^{-6})$    & $10.3^{+4.0}_{-2.8}$ & $14.0^{+6.6}_{-4.1}$
		&  $8.1^{+2.6}_{-2.0}$ &  [undeterm]   \\ [3pt]   
\hline \\ 
$\chi_\nu^2$~(dof) &   1.11~(30) & 1.22~(25)
		& 1.06~(30)   &   [undeterm]    \\ [3pt] 
$L_{\rm{0.3-8}}~(\times 10^{38}$~erg~s$^{-1})$ & $1.2^{+0.1}_{-0.1}$ & $1.6^{+1.8}_{-0.5}$
		& $1.1^{+0.1}_{-0.1}$ &  [undeterm]      \\[3pt]
\hline
\hline \\ 
\multicolumn{5}{c}{model: wabs$_{\rm Gal}$ $\times$ wabs $\times$ bremsstrahlung  }\\[5pt] 
\hline \\
$n_{\rm H}~(\times 10^{21}$~cm$^{-2})$  & $1.1^{+1.0}_{-0.9}$ & $0.8^{+0.7}_{-0.6}$ 
		 & $0.6^{+0.7}_{-0.6}$  &  $9.7^{+0.7}_{-0.8}$\\[3pt] 
$T_{\rm br}$~(keV)      & $10.8^{+17.5}_{-5.5}$ & $0.71^{+0.20}_{-0.19}$ 
		& $27.7^{+\ast}_{-18.6}$ &  $0.10^{+0.04}_{-0.03}$    \\[3pt]   
$K_{\rm br}~(\times 10^{-5})$    & $1.4^{+0.2}_{-0.2}$ & $2.8^{+0.5}_{-0.4}$
		&  $1.4^{+1.0}_{-0.3}$ & $1.0^{+0.4}_{-0.3}\,10^5$    \\ [3pt]   
\hline \\ 
$\chi_\nu^2$~(dof) &   1.05~(30) & 1.23~(25)
		& 1.05~(30)   &   1.55~(24)    \\ [3pt] 
$L_{\rm{0.3-8}}~(\times 10^{38}$~erg~s$^{-1})$ & $1.1^{+0.1}_{-0.1}$ & $0.6^{+0.2}_{-0.2}$
		& $1.1^{+0.1}_{-0.1}$ &  [undeterm]      \\[3pt]
\hline
\hline \\ 
\multicolumn{5}{c}{model: wabs$_{\rm Gal}$ $\times$ wabs $\times$ raymond-smith  }\\[5pt]  
\hline \\
$n_{\rm H}~(\times 10^{21}$~cm$^{-2})$  & $0.9^{+1.0}_{-0.7}$  & $0.9^{+0.5}_{-0.9}$ 
	& $0.7^{+0.6}_{-0.7}$ & $<0.5$ \\[3pt] 
$T_{\rm rs}$~(keV)       & $12.1^{+32.9}_{-5.3}$ & $0.65^{+0.40}_{-0.13}$ 
	& $24.6^{+39.4}_{-16.5}$ & $0.67^{+0.10}_{-0.08}$ \\[3pt]
Z (metal ab.) & 1 [fixed] & $<0.1$  
	& 1 [fixed] & $<0.2$  \\[3pt]
$K_{\rm rs}~(\times 10^{-5})$    & $2.6^{+2.2}_{-\ast}$ & $9.6^{+2.8}_{-\ast}$ 
	& $3.4^{+2.1}_{-\ast}$ & $2.9^{+2.1}_{-0.8}$\\ [3pt]   
\hline \\ 
$\chi_\nu^2$~(dof) &  1.04~(30) &  1.28~(24) & 1.05~(30) & 1.27~(23)\\ [3pt] 
$L_{\rm{0.3-8}}~(\times 10^{38}$~erg~s$^{-1})$ & $1.2^{+0.2}_{-0.2}$ & $0.6^{+0.2}_{-0.2}$ 
	& $1.1^{+0.2}_{-0.2}$ & $0.4^{+0.2}_{-0.2}$ \\[5pt]
\hline
\hline \\ 
\multicolumn{5}{c}{model: wabs$_{\rm Gal}$ $\times$ wabs $\times$ 
(powerlaw/bremss continuum $+$ gaussians) }\\[5pt] 
\hline \\
$n_{\rm H}~(\times 10^{21}$~cm$^{-2})$  & $0.2^{+0.8}_{-0.2}$ & $0.6^{+0.5}_{-0.5}$ 
		 & $0.5^{+0.6}_{-0.4}$  &  $9.6^{+0.6}_{-0.6}$\\[3pt] 
$\Gamma$       & $1.29^{+0.29}_{-0.20}$ &  
		& $1.30^{+0.15}_{-0.18}$ &       \\[3pt]   
$K_{\rm pl}~(\times 10^{-6})$    & $7.0^{+3.8}_{-2.5}$ & 
		&  $7.1^{+0.7}_{-0.9}$ &     \\ [5pt]   
$T_{\rm br}$~(keV)   &	& $0.72^{+0.16}_{-0.20}$	& 	& $0.15^{+0.05}_{-0.05}$\\[3pt]
$K_{\rm br}~(\times 10^{-5})$   &	&  $2.5^{+1.9}_{-1.1}$	& & $526^{+60}_{-60}$\\[3pt]
line 1 	& $E = 1.32^{+0.04}_{-0.05}$ keV  & $E = 1.27^{+0.03}_{-0.03}$ keV &
	$E = 1.33^{+0.08}_{-0.08}$	& \\[3pt]
	& $EW = 85^{+108}_{-60}$ eV	& $EW = 176^{+103}_{-84}$ eV	 
	& $EW = 73^{+94}_{-73}$ eV	&  \\ [3pt]
line 2 	& $E = 1.51^{+0.04}_{-0.03}$ keV & 	& 	& $E = 1.50^{+0.04}_{-0.05}$ keV \\[3pt]
	& $EW = 135^{+89}_{-88}$ eV	&	&	&  [undeterm]\\ [3pt]
line 3 	& $E = 1.85^{+0.04}_{-0.03}$ keV & 	& $E = 1.91^{+0.04}_{-0.04}$ keV 	
	& $E = 1.89^{+0.03}_{-0.07}$ keV \\[3pt]
	& $EW = 154^{+98}_{-131}$ eV	&	& $EW = 189^{+114}_{-111}$ eV	
	& [undeterm]\\ [3pt]
line 4 	& $E = 2.60^{+0.19}_{-0.10}$ keV & 	& 	& \\[3pt]
	& $EW = 293^{+293}_{-204}$ eV	&	&	&  \\ [3pt]
\hline \\ 
$\chi_\nu^2$~(dof) &   0.82~(22) & 0.88~(23)
		& 0.85~(24)   &  1.01~(22)     \\ [3pt] 
$L_{\rm{0.3-8}}~(\times 10^{38}$~erg~s$^{-1})$ & $1.1^{+0.1}_{-0.1}$ & $0.5^{+0.2}_{-0.1}$
		& $1.1^{+0.1}_{-0.1}$ &  [undeterm]      \\[3pt]
\hline \\
\end{tabular} 
\end{table*}



\begin{table*}
\caption{
   {\small XSPEC} best-fit parameters 
   for other selected bright sources in the ACIS-S3 chip.
(An asterisk indicates 
that we could not obtain a meaningful best-fit value for that parameter).}
\centering 
\begin{tabular}{@{}lrrrrrrrr}
\hline
\hline \\ 
parameter &  No.~5 & No.~31 & No.~33  & [No.~39] & No.~44 & No.~60 & No.~64 & No.~72 \\  [5pt] 
\hline
\hline \\ 
\multicolumn{9}{c}{model: wabs$_{\rm Gal}$ $\times$ wabs $\times$ powerlaw  }\\[5pt] 
\hline \\
$n_{\rm H}~(\times 10^{21}$~cm$^{-2})$  & $3.5^{+2.0}_{-1.3}$ & $1.05^{+0.84}_{-0.94}$ 
		 & $1.15^{+0.47}_{-0.51}$  &$7.7^{+7.5}_{-3.1}$   & $0.74^{+0.26}_{-0.29}$ 
		& $1.13^{+0.48}_{-0.77}$ 
		& $2.1^{+0.4}_{-0.4}$ & $8.0^{+1.7}_{-1.4}$ \\[5pt] 
$\Gamma$       & $1.62^{+0.33}_{-0.31}$ & $1.39^{+0.21}_{-0.32}$ 
		& $1.68^{+0.16}_{-0.17}$ &   $1.43^{+0.51}_{-0.40}$   & $1.40^{+0.12}_{-0.10}$ 
		& $1.58^{+0.22}_{-0.24}$
		& $2.77^{+0.21}_{-0.18}$ & $3.13^{+0.57}_{-0.11}$  \\[5pt]   
$K_{\rm pl}~(\times 10^{-5})$    & $1.6^{+0.6}_{-0.6}$ & $0.85^{+0.36}_{-0.25}$
		& $1.3^{+0.3}_{-0.3}$ &    $0.59^{+0.10}_{-0.09}$      & $2.7^{+0.4}_{-0.3}$  
		& $1.0^{+0.2}_{-0.2}$ 
		& $6.3^{+1.0}_{-0.9}$ & $7.4^{+3.3}_{-2.1}$  \\ [5pt]   
\hline \\ 
$\chi_\nu^2$~(dof) &  0.88~(33) & 0.88~(27)
		&  1.08~(41)  &    0.72~(9)   &  1.24~(67) &  0.60~(34) 
		&  0.94~(94) &  1.58~(27)  \\ [5pt] 
$L_{\rm{0.3-8}}~(\times 10^{38}$~erg~s$^{-1})$ & $1.9^{+0.2}_{-0.4}$ & $1.1^{+0.1}_{-0.1}$
		& $1.3^{+0.1}_{-0.1}$ &   [Radio gal.]       &$3.6^{+0.1}_{-0.1}$ & $1.1^{+0.1}_{-0.1}$ 
		& $5.0^{+1.1}_{-0.8}$ & $7.1^{+4.8}_{-2.5}$ \\[5pt]
\hline
\hline \\ 
\multicolumn{9}{c}{model: wabs$_{\rm Gal}$ $\times$ wabs $\times$ disk-blackbody  }\\[5pt] 
\hline \\ 
$n_{\rm H}~(\times 10^{21}$~cm$^{-2})$  &&&
		&&&& $0.13^{+0.22}_{-0.13}$
		& $3.4^{+1.1}_{-1.0}$  \\[5pt] 
$T_{\rm in}$~(keV)    &&&
		&&&& $0.68^{+0.07}_{-0.07}$
  		& $0.72^{+0.09}_{-0.08}$  \\[5pt]
$K_{\rm dbb}~(\times 10^{-2})$    &&&
		&&&& $2.9^{+1.7}_{-1.0}$
  		& $1.6^{+1.3}_{-0.7}$ \\ [5pt]   
\hline \\ 
$\chi_\nu^2$~(dof) &&&
		&&&& 1.13~(94) 
		&  1.21~(27)  \\ [5pt] 
$L_{\rm{0.3-8}}~(\times 10^{38}$~erg~s$^{-1})$ &&&
		&&&& $2.0^{+0.1}_{-0.1}$
		& $1.4^{+0.3}_{-0.1}$  \\[5pt]
\hline
\hline \\ 
\multicolumn{9}{c}{model: wabs$_{\rm Gal}$ $\times$ wabs $\times$ bmc  }\\[5pt] 
\hline \\
$n_{\rm H}~(\times 10^{21}$~cm$^{-2})$   &&& 
		&&&& $0.65^{+0.61}_{-0.64}$
		& $1.1^{+1.5}_{-0.9}$\\[5pt] 
$T_{\rm bb}$~(keV)   &&& 
		&&&& $0.18^{+0.06}_{-0.10}$
		& $0.53^{+0.04}_{-0.12}$\\[5pt] 
$\Gamma$         &&&
		&&&& $2.56^{+0.19}_{-0.23}$
		& $6.3^{+\ast}_{-\ast}$\\[5pt]
$K_{\rm bmc}~(\times 10^{-7})$        &&&
		&&&& $11.9^{+0.4}_{-0.4}$
		& $7.1^{+0.8}_{-0.6}$\\ [5pt]   
\hline \\  
 $\chi_\nu^2$~(dof)   &&&
		&&&& 0.90~(92)
		& 1.28~(25)  \\ [5pt] 
$L_{\rm{0.3-8}}~(\times 10^{38}$~erg~s$^{-1})$  &&&
		&&&& $2.5^{+0.6}_{-0.3}$
		& $1.0^{+0.1}_{-0.1}$ \\[5pt] 
\hline \\
\end{tabular}  
\end{table*}



\begin{table*}
\centering 
\begin{tabular}{@{}lrrrrrrrr}
\hline
\hline \\
parameter &  No.~73 & No.~78 & No.~85 & No.~86 & No.~88 & No.~104 & No.~113 & No.~121\\  [5pt]    
\hline
\hline \\ 
\multicolumn{9}{c}{model: wabs$_{\rm Gal}$ $\times$ wabs $\times$ powerlaw  }\\[5pt]
\hline \\
$n_{\rm H}~(\times 10^{21}$~cm$^{-2})$  & $1.8^{+0.6}_{-0.6}$ & $<0.14$ 
		& $0.84^{+0.54}_{-0.57}$ & $3.9^{+0.5}_{-0.4}$ & $2.9^{+0.4}_{-0.4}$
		&  $14.6^{+4.5}_{-2.5}$ & $16.7^{+8.6}_{-6.8}$ & $1.17^{+0.79}_{-0.85}$ \\[5pt] 
$\Gamma$       &$2.30^{+0.26}_{-0.19}$ & $0.98^{+0.20}_{-0.18}$ 
		& $1.65^{+0.27}_{-0.23}$ &  $2.69^{+0.15}_{-0.15}$ & $2.37^{+0.08}_{-0.14}$
		& $2.12^{+0.32}_{-0.27}$& $2.17^{+0.74}_{-0.44}$ & $1.29^{+0.15}_{-0.17}$ \\[5pt]   
$K_{\rm pl}~(\times 10^{-5})$    &$2.0^{+0.6}_{-0.4}$ & $0.51^{+0.06}_{-0.07}$ 
		&  $0.85^{+0.23}_{-0.20}$ & $9.2^{+1.3}_{-1.2}$ & $7.5^{+1.1}_{-0.9}$ 
		& $9.3^{+5.2}_{-3.2}$&  $2.4^{+0.4}_{-0.3}$ & $1.2^{+0.4}_{-0.2}$ \\ [5pt]   
\hline \\ 
$\chi_\nu^2$~(dof) &1.09~(41) & 1.17~(29) 
		&  0.86~(30) &   0.75~(105) & 0.80~(105)
		&  0.95~(63) & 0.53~(14) &  0.79~(46)\\ [5pt] 
$L_{\rm{0.3-8}}~(\times 10^{38}$~erg~s$^{-1})$ &$1.6^{+0.2}_{-0.3}$ &  $1.0^{+0.1}_{-0.1}$ 
		& $0.9^{+0.1}_{-0.1}$ &  $7.2^{+1.4}_{-0.9}$ &  $5.9^{+0.8}_{-0.7}$
		& $7.7^{+3.5}_{-1.5}$& $2.0^{+2.8}_{-0.8}$ & $1.8^{+0.1}_{-0.1}$\\[5pt]
\hline
\hline \\ 
\multicolumn{9}{c}{model: wabs$_{\rm Gal}$ $\times$ wabs $\times$ disk-blackbody  }\\[5pt] 
\hline \\ 
$n_{\rm H}~(\times 10^{21}$~cm$^{-2})$      & $<0.39$ & &  & $1.32^{+0.29}_{-0.26}$
		&  $0.90^{+0.24}_{-0.23}$  
  		&  $8.9^{+2.4}_{-2.1}$ & $10.2^{+5.8}_{-3.3}$&\\[5pt] 
$T_{\rm in}$~(keV)    & $0.94^{+0.15}_{-0.16}$ & & & $0.86^{+0.07}_{-0.06}$ 
		&  $0.91^{+0.08}_{-0.07}$  
  		&  $1.46^{+0.27}_{-0.21}$ & $1.50^{+0.69}_{-0.44}$&\\[5pt]
$K_{\rm dbb}~(\times 10^{-2})$    & $0.33^{+0.46}_{-0.13}$ &&  & $1.7^{+0.6}_{-0.5}$
		&  $1.4^{+0.5}_{-0.4}$ 
  		&  $0.27^{+0.12}_{-0.09}$ & $0.06^{+0.21}_{-\ast}$ &\\ [5pt]   
\hline \\ 
$\chi_\nu^2$~(dof) & 1.42~(41)& & & 0.68~(105)
		&   0.81~(112)  
		&   0.87~(63)&  0.54~(14)&\\ [5pt] 
$L_{\rm{0.3-8}}~(\times 10^{38}$~erg~s$^{-1})$ & $0.9^{+0.1}_{-0.1}$ & & & $3.0^{+0.1}_{-0.1}$
		&  $3.0^{+0.1}_{-0.1}$  
		&   $4.0^{+0.3}_{-0.3}$& $1.0^{+0.2}_{-0.1}$&\\[5pt]
\hline
\hline \\ 
\multicolumn{9}{c}{model: wabs$_{\rm Gal}$ $\times$ wabs $\times$ bmc  }\\[5pt] 
\hline \\
$n_{\rm H}~(\times 10^{21}$~cm$^{-2})$  &$14.9^{+19.7}_{-12.4}$ &   
		& & $0.8^{+0.3}_{-0.4}$ & $0.55^{+0.24}_{-0.20}$
		&  $5.1^{+3.6}_{-2.9}$ & $9.1^{+17.5}_{-3.3}$   \\[5pt] 
$T_{\rm bb}$~(keV)   & $0.19^{+0.08}_{-0.07}$ &  
		& & $0.36^{+0.07}_{-0.04}$&  $0.31^{+0.07}_{-0.02}$
		&  $0.76^{+0.19}_{-0.31}$& $0.60^{+0.36}_{-\ast}$ \\[5pt] 
$\Gamma$         &$1.67^{+0.54}_{-0.58}$ & 
		& & $3.25^{+0.29}_{-0.34}$& $2.61^{+0.28}_{-0.21}$
		&  $1.3^{+2.7}_{-0.2}$ & $\ast$ \\[5pt]
$K_{\rm bmc}~(\times 10^{-7})$        & $6.7^{+0.3}_{-0.3}$ & 
		& & $14.6^{+5.3}_{-0.8}$& $14.3^{+1.2}_{-0.7}$
		&  $3.0^{+0.6}_{-0.6}$ & $7.6^{+\ast}_{-\ast}$ &   \\ [5pt]   
\hline \\  
 $\chi_\nu^2$~(dof)   & 0.93~(39)& 
		& & 0.72~(103)& 0.79~(110)
		&   0.89~(61) &  0.54~(12)   \\ [5pt] 
$L_{\rm{0.3-8}}~(\times 10^{38}$~erg~s$^{-1})$  & $1.4^{+1.5}_{-0.3}$ & 
		& & $2.7^{+0.1}_{-0.1}$& $3.1^{+0.1}_{-0.1}$
		&  $3.3^{+0.4}_{-0.4}$ & $0.9^{+\ast}_{-\ast}$\\[5pt] 
\hline \\
\end{tabular}  
\end{table*}

\clearpage 

\begin{figure*}
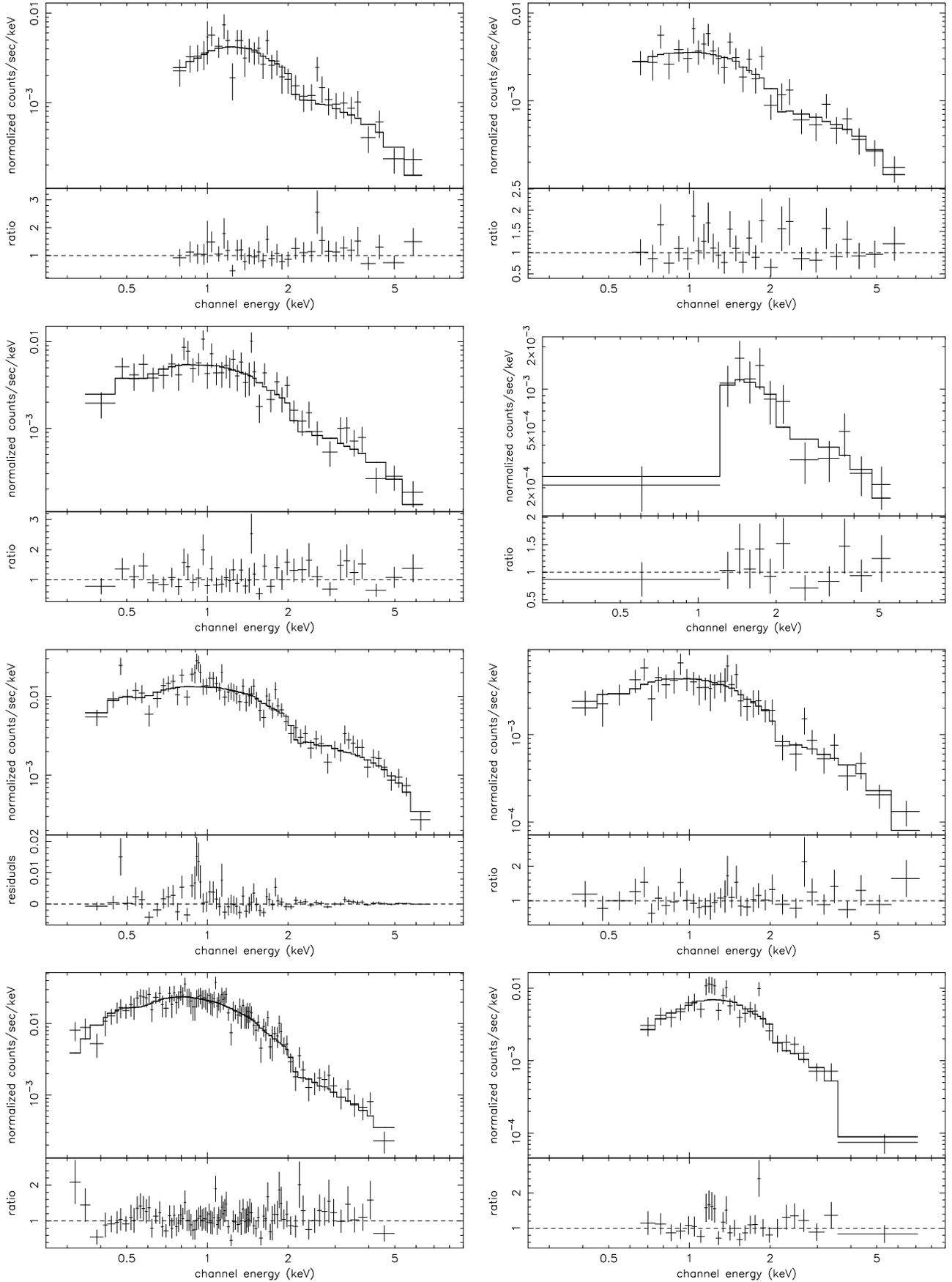

\vbox{
\begin{tabular}{lr}
\psfig{figure=s05.ps,width=5.6cm, angle=270} &
\psfig{figure=s31.ps,width=5.6cm, angle=270} \\
\psfig{figure=s33.ps,width=5.6cm, angle=270} &
\psfig{figure=s39.ps,width=5.6cm, angle=270} \\
\psfig{figure=s44.ps,width=5.6cm, angle=270} &
\psfig{figure=s60.ps,width=5.6cm, angle=270} \\ 
\psfig{figure=s64.ps,width=5.6cm, angle=270} &
\psfig{figure=s72.ps,width=5.6cm, angle=270} 
\end{tabular}
}
\caption{Spectra of bright sources of group C: from top left: 
  No.~5 and 31; 
  No.~33 (a candidate X-ray pulsar, lightcurve shown in Fig.~9) 
  and 39 (a background radio galaxy); 
  No.~44 and 60; 
   No.~64 and 72.    
Their fitting parameters are listed in Table B.4. 
}
\label{fig:imagea1}
\end{figure*}


                             
\begin{figure*}
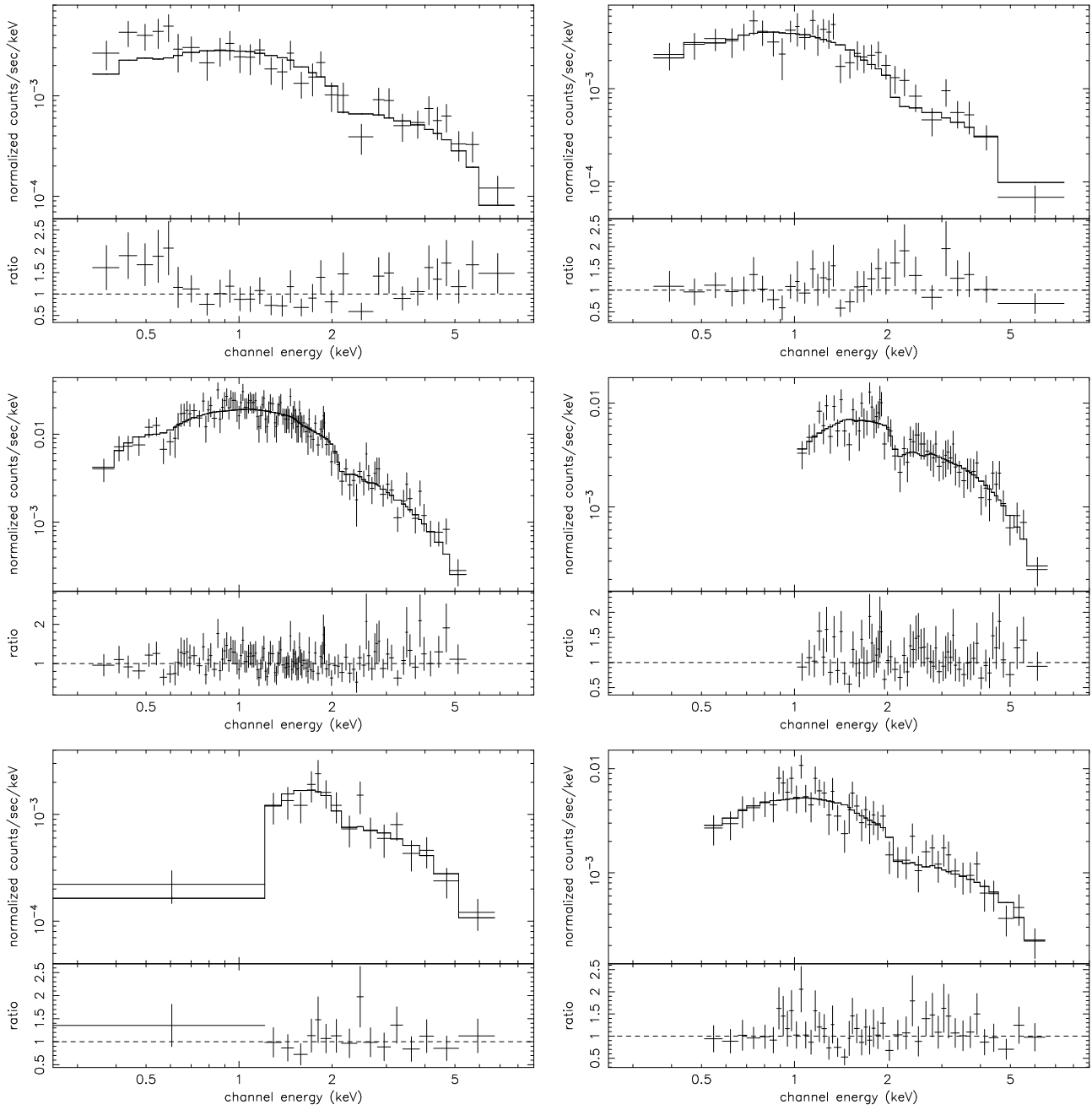

\vbox{
\begin{tabular}{lr}
\psfig{figure=s78.ps,width=5.6cm, angle=270} &
\psfig{figure=s85.ps,width=5.6cm, angle=270} \\
\psfig{figure=s88.ps,width=5.6cm, angle=270} &
\psfig{figure=s104.ps,width=5.6cm, angle=270} \\ 
\psfig{figure=s113.ps,width=5.6cm, angle=270} &
\psfig{figure=s121.ps,width=5.6cm, angle=270} \\
\end{tabular}
}
\caption{Spectra of bright sources of group C (continued): 
  from top left: 
  No.~78 and 85; 
  No.~88 and 104.  
  No.~113 (a candidate X-ray pulsar, lightcurve shown in Fig.~10) and No.~121.
  Their fitting parameters are listed in Table B.4.
}
\label{fig:imagea2}
\end{figure*}



\begin{thebibliography}{}    



  \bibitem[1996]{Arnaud}
	Arnaud, K. A. 1996, Astronomical Data Analysis Software 
	and Systems V, eds. G. Jacoby and J. Barnes, ASP Conference Series 
	Volume 101, p.~17

  \bibitem[1997]{Blair} 
      Blair, W. P., \& Long, K. S. 
      1997, ApJS, 108, 261 
 
  \bibitem[1983]{Blundell}
      Blundell, K. M., Rawlings, S., \& Willott, C. J. 
      1999, AJ, 117, 677

  \bibitem[1983]{Bohlin}  
      Bohlin, R. C., Cornett, R. H., Hill, J. K., Smith, A. M., \& Stecher, T. P., 
      1983, ApJ, 274, L53  

    \bibitem[2002]{Buat} 
      Buat, V., Boselli, A., Gavazzi, G., \& Bonfanti, C. 
      2002, A\&A 383, 801  

    \bibitem[1994]{Chevalier}
	Chevalier, R. A., \& Fransson, C. 1994, ApJ, 420, 268

    \bibitem[1996]{Corbet1}
	Corbet, R. H. D. 1986, MNRAS, 220, 1047

    \bibitem[1996]{Corbet2}
	Corbet, R. H. D., Marshall, F. E., Coe, M. J., Laycock, S., \& Handler, G.
	2001, ApJ, 548, 41L

  \bibitem[1998]{Cote} 
      C{\^ o}t{\' e}, S., Freeman, K. C., Carignan, C., \& Quinn, P. 
      1997, AJ, 114, 1313  

  \bibitem[1985]{Cowan}      
      Cowan, J. J., \& Branch, D. 
      1985, ApJ, 293, 400

  \bibitem[1994]{Cowan2}      
      Cowan, J. J., Roberts, D. A., \& Branch, D. 
      1994, ApJ, 434, 128

  \bibitem[1979]{deVaucouleurs1}   
      de Vaucouleurs, G. 
      1979, AJ, 84, 1270  
      
  \bibitem[1983]{deVaucouleurs2}   
      de Vaucouleurs, G., Pence, W. D., \& Davoust, E.  
      1983, ApJS, 53, 17      

  \bibitem[1991]{deVaucouleurs3}   
      de Vaucouleurs, G., de Vaucouleurs, A., Corwin, H. Jr., et al.   
      1991, Third Reference Catalogue of Bright Galaxies,  
        (Springer-Verlag, New York) 

  \bibitem[2002]{DiStefano} 
      Di Stefano, R., \& Kong, A. K. H. 
      2003, ApJ, in press (astro-ph/0301162)
   
            
  \bibitem[1998]{Ehle}     
      Ehle, M., Pietsch, W., Beck, R., \& Klein, U.  
      1998, A\&A, 329, 39    
      
  \bibitem[1998]{Elmegreen} 
      Elmegreen, D. M., Chromey, F. R., \& Warren, A. R. 
      1998, AJ, 116, 2834  

  \bibitem[1984]{Fabb}
      Fabbiano, G., Trinchieri, G., Elvis, M.,
 	Miller, L., \& Longair, M. 
      1984, ApJ, 277, 115

  \bibitem[2001]{Fabbiano}
      Fabbiano, G., Zezas, A., \& Murray, S. S. 
      2001, ApJ, 554, 1035  

  \bibitem[1991]{Gallais} 
      Gallais, P., Rouan, D., Lacombe, F., Tiphe\`ene, D., \& Vauglin, I. 
      1991, A\&A, 243, 309 
      
  \bibitem[2001]{Harris}  
      Harris, J., Calizetti, D., Gallagher, J. S. III, Conselice, C. J., 
         \& Smith, D. A. 
      2001, AJ, 122, 3046      
      
  \bibitem[1997]{Heckman} 
      Heckman, T. M. 
      1997, in the proceedings of the conference ``Origins'', 
	Ed. C. Woodward and J.M. Shull (PASP), astro-ph/9708263   

  \bibitem[1994]{Hughes}   
      Hughes, J. P. 
      1994, ApJ, 427, L25
      
  \bibitem[1999]{Immler}     
      Immler, S., Vogler, A., Ehle, M., \& Pietsch, W. 
      1999, A\&A, 352, 415     

  \bibitem[1999]{Kahabka}    
      Kahabka, P.,  
      1999, A\&A, 344, 459     
  
  \bibitem[1999]{Kembhavi}
      Kembhavi, A. K., \& Narlikar, J. V.
      1999, in Quasars and Active Galactic Nuclei, 
	(Cambridge University Press: Cambridge)  

  \bibitem[2002]{Kilgard} 
	Kilgard, R. E., Kaaret, P., Krauss, M. I., et al.~2002, ApJ, 570, 671

  \bibitem[2003]{Kong} 
      Kong, A. K. H., Di Stefano, R., Garcia, M., \& Greiner, J. 
      2003, ApJ, 585, 298

  \bibitem[2001]{Labarbera}
	La Barbera, A., Burderi, L., Di Salvo, T., Iaria, R., \& Robba, N. R.
  	2001, ApJ, 553, 375

  \bibitem[2002]{Laycock}
	Laycock, S., Corbet, R. H. D., Perrodin, D., et al.~2002, A\&A, 385, 464

  \bibitem[1998]{Lewin} 
  	Lewin, W. H. G., van Paradijs, J., \& van den Heuvel, E. P. J. 1995, 
	X-ray binaries (Cambridge University Press: Cambridge) 

  \bibitem[1998]{Ma}    
      Ma, C., Arias, E. F., Eubanks, T. M., et al.~1998, AJ, 116, 516

  \bibitem[2001]{Matsumoto}
      Matsumoto, H., Tsuru, T. G., Koyama, K., Awaki, H., Canizares, C. R., 
	Kawai, N., Matsushita, S., \& Kawabe, R.
      2001, ApJ, 547, L25  

  \bibitem[1993]{Mavro}
      Mavromatakis, F. 
      1993, A\&A, 273, 147 

  \bibitem[1998]{Negueruela}
	Negueruela, I. 1998, A\&A, 338, 505

   
  \bibitem[1990]{Ohashi} 
      Ohashi, T., Makishima, K., Tsuru, T., et al.~1990, ApJ, 365, 180
            
  \bibitem[1997]{Okada}     
      Okada, K., Mitsuda, K., \& Dotani, T.   
      1997, PASJ, 49, 653  
      
  \bibitem[1994]{Owen}
      Owen, F. N., \& Ledlow, M. J. 
      1994, proceedings of ``The First Stromlo Symposium: The Physics of Active Galaxies'', 
	ASP Conference Series, Vol. 54, eds G. V. Bicknell, M. A.
        Dopita, and P. J. Quinn, p.319 

  \bibitem[2002]{Pannuti}  
      Pannuti, T. G. 2002, PhD thesis (available from 
	http://space.mit.edu/$\sim$tpannuti)

  \bibitem[2002]{Pannuti}  
      Pannuti, T. G., Duric, N., Lacey, C. K., et al.~2002, ApJ, 565, 966

  \bibitem[2002]{Pooley}
	Pooley, D., et al. 2002, ApJ, 572, 932

  \bibitem[1995]{Predehl} 
      Predehl, P., \& Schmitt, J. H. M. M. 
      1995, A\&A, 293, 889    
      
  \bibitem[2002]{Prestwich1}
      Prestwich, A. H. 
      2001, proceedings of ``X-rays at Sharp Focus: 
	{\it Chandra} Science Symposium'', 
        eds. S. Vrtilek, E. M. Schlegel, \& E. Kuhi  
        (astro-ph/0107133)    

  \bibitem[2003]{Prestwich2}
      Prestwich, A. H., Irwin, J. A., Kilgard, R. E., et al.~2003, 
	ApJ, in press (astro-ph/0206127)

  \bibitem[1994]{Rappaport}  
      Rappaport, S., Di Stefano, R., \& Smith, J. D. 
      1994, ApJ, 426, 692     

  \bibitem[2002]{Rosati}    
      Rosati, P., et al. 
      2002, ApJ, 566, 667  
      
  \bibitem[1999]{Sako}
	Sako, M., Liedahl, D. A., Kahn, S. M., \& Paerels, F. 1999, 
	ApJ, 525, 921

  \bibitem[1987]{Sandage} 
      Sandage, A., \& Tammann, G. A.  
      1987, A revised Shapley-Ames Catalog of Bright Galaxies, 
        2nd ed. (Carnegie Institution of Washington Publication: 
        Washington)   
         
  \bibitem[1998]{Schelegel}
      Schlegel, D. E., Finkbeiner, D. P., \& Davis, M. 
      1998, ApJ, 500, 525 
               
  \bibitem[2001]{Shirey} 
      Shirey, R., Soria, R., Borozdin, K., et al. 2001, A\&A, 365, L195              

  \bibitem[2002]{Soria2} 
      Soria, R. 2002, proceedings of "High Energy Processes 
	and Phenomena in Astrophysics", IAU Symposium No.214, Suzhou
     	(China), 5-10 August 2002; eds.: X. Li, Z. Wang, V. Trimble 
	(astro-ph/0211024)
      
  \bibitem[2002]{Soria1} 
      Soria, R., \& Wu, K. 
      2002, A\&A, 384, 99  

  \bibitem[2001]{Stockdale}
      Stockdale, C. J., Cowan, J. J., Maddox, L. A., et al.~2001, BAAS, 199, 1902

  \bibitem[2001]{Sunyaev1}
	Sunyaev, R. 2001, Proceedings of the ESO Workshop (Garching, Sept. 1999) 
	eds. L. Kaper et al. (Springer), astro-ph/0103469

  \bibitem[2001]{Sunyaev2}
	Sunyaev, R., \& Revnivtsev, M. 2000,
 	A\&A, 358, 617

  \bibitem[2001]{Supper} 
      Supper, R., Hasinger, G., Lewin, W. H. G., et al.~2001, A\&A, 373, 63
      
  \bibitem[2003]{Swartz1} 
      Swartz, D. A., Ghosh, K. K., McCollough, M. L., et al.~2003, ApJS, 144, 213

  \bibitem[2002]{Swartz2} 
      Swartz, D. A., Ghosh, K. K., Suleimanov, V., Tennant, A. F., \& Wu, K.
      2002, ApJ, 574, 382
       
  \bibitem[1979]{Talbot} 
      Talbot, R. J., Jensen, E. B., \& Dufour, R. J.
      1979, ApJ, 229, 91    

  \bibitem[2001]{Tennant}
      Tennant, A. F., Wu, K., Ghosh, K. K., Kolodziejcak, J. J., \& 
        Swartz, D. A. 
      2001, ApJ, 549, L43    

  \bibitem[1968]{Thackeray}   
      Thackeray, A. D. 
      1968, Sky \& Telescope, 36, 295 

  \bibitem[2000]{Thatte} 
      Thatte, N., Tecza, M., \& Genzel, R. 
      2000, A\&A, 364, L47      

  \bibitem[1993]{Tilanus}  
      Tilanus, R. P. J., \& Allen, R. J. 
      1993, A\&A, 274, 707 
            
  \bibitem[1985]{Trinchieri}     
      Trinchieri, G., Fabbiano, G., \& Palumbo, C. G. C.   
      1985, ApJ, 290, 96    

  \bibitem[1980]{vandenBergh} 
      van den Bergh, S. 
      1980, PASP, 92, 122  	 

  \bibitem[1992]{vandenHeuvel}   
      van den Heuvel, E. P. J., Bhattacharya, D., Nomoto, K., \& Rappaport, S. A. 
      1992, A\&A, 262, 97    
         

  \bibitem[1974]{wood} 
      Wood, R., \& Andrews, P. J. 
      1974, MNRAS, 167, 13 

  \bibitem[2001]{wu1} 
      Wu, K. 
      2001, Publ. Astron. Soc. Australia, 18, 443  

  \bibitem[2003]{wu2} 
      Wu, K., Tennant, A. F., Swartz, D. A., Ghosh, K. K., \& Hunstead, R. W.~2003,
	proceedings of the symposium 
	``New Visions of the X-ray Universe in the XMM-Newton and Chandra Era'',
  	ESTEC, The Netherlands, Nov 2001 (astro-ph/0302363)  
  

  \bibitem[2002]{zezas} 
      Zezas, A., Fabbiano, G., Rots, A. H., \& Murray, S. S. 
      2002, ApJS, 142, 239

\end{thebibliography}
\end{document}